\documentclass[11pt]{amsart}

\usepackage{graphicx, amsmath, amssymb, latexsym, amsfonts, amsthm, amssymb,epstopdf,lscape, longtable}

\newcommand{\comment}[1]{}

\newcommand{\be}{\begin{enumerate}}
\newcommand{\ee}{\end{enumerate}}

\newtheorem{theorem}{Theorem}[section]

\newtheorem{definition}[theorem]{Definition}

\begin{document}

\title{Public Key Exchange Using Matrices Over Group Rings}
\author[D. Kahrobaei]{Delaram Kahrobaei}
\address{CUNY Graduate Center and City Tech, City University of New York}%
\email{DKahrobaei@GC.Cuny.edu}
\thanks{Research of the first author was partially supported by a PSC-CUNY grant from
the CUNY research foundation, as well as the City Tech foundation.}
\author[C. Koupparis]{Charalambos Koupparis}
\address{CUNY Graduate Center, City University of New York}%
\email{ckoupparis@GC.Cuny.edu}
\author{Vladimir Shpilrain}
\address{The City College of New York and CUNY Graduate Center}
\email{shpil@groups.sci.ccny.cuny.edu}
\thanks{Research of the third author was partially supported by
the NSF grants DMS-0914778 and CNS-1117675.}

\begin{abstract}

We offer a public key exchange protocol in the spirit of
Diffie-Hellman, but we use (small) matrices over a group ring of a
(small) symmetric group as the platform. This ``nested structure" of
the platform makes computation very efficient for legitimate
parties. We discuss security of this scheme by addressing the
Decision Diffie-Hellman (DDH)  and Computational Diffie-Hellman
(CDH) problems for our platform.


\end{abstract}

\maketitle

\pagenumbering{arabic}

\section{Introduction}

The beginning of public key cryptography can be traced back to the
paper  by Diffie and Hellman \cite{DH}. The simplest, and original,
implementation of their  key exchange protocol uses
$\mathbb{Z}_p^\ast$, the multiplicative group of integers modulo a
prime $p$, as the platform. There is also a public element $g \in
\mathbb{Z}_p$, which is a primitive root mod $p$. The protocol
itself is as follows:

\be
\item Alice chooses an integer $a$,
computes $A=g^a \text{ mod } p $ and publishes $A$
\item Bob picks an integer $b$ and computes $B=g^b \text{ mod } p $, and publishes $B$
\item Alice computes $K_A = B^a \text{ mod } p$
\item Bob computes $K_B = A^b\text{ mod } p$
\ee

Both Alice and Bob are now in possession of a secret shared key $K$,
as $g^{ab} \text{ mod } p  = g^{ba} \text{ mod } p$ and hence
$K:=K_A=K_B$.

The protocol is considered secure provided $G$ and $g$ are chosen
properly,  see e.g. \cite{Menezes} for details. In order to recover
the shared secret key, the eavesdropper Eve must be able to solve
the \textit{Diffie-Hellman problem} (recover $g^{ab}$ from $g, g^a$
and $g^b$). One could solve the Diffie-Hellman problem by  solving
the discrete logarithm problem, i.e., by recovering $a$ from $g$ and
$g^a$. However, it is unknown whether the discrete logarithm problem
is equivalent to the Diffie-Hellman problem.

We should note that there is still the ``brute force'' method of
solving the  discrete logarithm problem. The eavesdropper can simply
start computing successively higher powers of $g$, until they match
$g^a$. This requires at most $|g|$ multiplications, where $|g|$ is
the order of $g$ in the group $G$. It is usually the case however
that $|g| \approx 10^{300}$ and hence this method is considered
computationally infeasible.

Initially it may seem that the legitimate parties, Alice and Bob,
will also have  to perform a large number of multiplications, thus
facing the same problem as the eavesdropper does. However, as the
legitimate parties are in possession of $a$ and $b$, they can use
the ``square and multiply" algorithm that requires  $O(\log_2 a)$
multiplications, e.g. $g^{27} = (((g^2)^2)^2)^2\cdot ((g^2)^2)^2
\cdot g^2 \cdot g$.

There is some disadvantage to working with $\mathbb{Z}_p$, where
$p$, $a$, and $b$ are chosen to be fairly large. Computation with
300-digit  numbers (or 1000-bit binary numbers) is not particularly
efficient, and neither is reducing the result modulo $p$. This is
one of the reasons why the Diffie-Hellman key agreement protocol
with recommended parameters is not suitable for devices with limited
computational resources. Hence, there is an ongoing search for other
platforms where the Diffie-Hellman or a similar key exchange can be
carried out more efficiently, in particular with public and/or
private keys of smaller size.

The platform that we are  proposing here is the semigroup of
matrices (of a small size) over a group ring, with the usual matrix
multiplication operation. More specifically, we are working with
matrices   over the group ring $\mathbb{Z}_n[S_m]$, where
$\mathbb{Z}_n$ is the ring of integers modulo $n$ and $S_m$ is the
symmetric group of degree $m$. To verify the security of using such
a semigroup of matrices as  the platform, we address the
\textit{Computational Diffie-Hellman} and \textit{Decision
Diffie-Hellman} problems (Section  \ref{DDH}), along with questions
about the structure of this semigroup.

Parameters that we suggest ($2\times 2$ or $3\times 3$ matrices over
$\mathbb{Z}_7[S_5]$) provide for a large key space ($7^{480}\sim
10^{406}$ for $2\times 2$ matrices and  $7^{1080}\sim 10^{913}$ for
$3\times 3$ matrices). Storing a single $2\times 2$  matrix over
$\mathbb{Z}_7[S_5]$ takes about 1440  bits, and a single $3\times 3$
matrix about 3240 bits, so keys are of about the same size as in the
``classical" Diffie-Hellman scheme (storing an integer of size about
$10^{300}$ requires 997 bits).  These storage requirements can be
reduced by $\frac{1}{7}{th}$ if we do not store polynomial terms
which have a $0$ as their coefficient, thus bringing the key size
down to about 1230 bits for $2\times 2$ matrices and to about 2780
bits for $3\times 3$ matrices.

What we believe is one of the main advantages of our  platform  over the
standard $\mathbb{Z}_p$  platform in the original  Diffie-Hellman
scheme is that the multiplication of matrices over
$\mathbb{Z}_7[S_5]$ is very efficient. In particular, in our setup
multiplying elements is faster than multiplying numbers in
$\mathbb{Z}_p$ for a large $p$. This is due to the fact that one can
pre-compute the multiplication table for the group $S_5$ (of order
120), so  in order to multiply two elements of $\mathbb{Z}_7[S_5]$
there is no ``actual" multiplication in $S_5$ involved, but just
re-arranging a bit string and multiplying coefficients in
$\mathbb{Z}_7$. Also, in our multiplication there is no reduction of
the result modulo $p$ that slows down computation in $\mathbb{Z}_p$
for a large $p$. Informally speaking, the ``nested structure"  of
our platform ({\it small} matrices over a group ring of a {\it
small} group $S_5$ over a {\it small}  ring $\mathbb{Z}_7$) provide
for more efficient computation than just using $\mathbb{Z}_p$ with a
very large  $p$.

From a security standpoint, an advantage of our platform over the
group  $\mathbb{Z}_p$, or elliptic curves, is that ``standard"
attacks (baby--step giant--step, Pohlig-Hellman,  Pollard's rho) do
not work with our platform, as we show in Section  \ref{attacks}.
Furthermore, our platform proves secure against Shor's quantum
algorithm which is a common pitfall on classical Diffie-Hellman
algorithms, see  Section \ref{quant}.



\section{Group Rings}

\begin{definition}

Let $G$ be a group written multiplicatively and let $R$ be any
commutative  ring with nonzero unity. The group ring $R[G]$ is
defined to be the set of all formal sums

\[\sum_{g_i \in G} r_i g_i\]
where $r_i \in R$, and all but a finite number of $r_i$ are zero.

\end{definition}

We define the sum of two elements in $R[G]$ by

\[\left(\sum_{g_i\in G}a_ig_i\right)+\left(\sum_{g_i\in G}b_ig_i\right) = \sum_{g_i \in G}(a_i+b_i)g_i.\]

Note that $(a_i+b_i)=0$ for all but a finite number of $i$, hence
the above sum is in $R[G]$.  Thus $(R[G],+)$ is an abelian group.

Multiplication of two elements of $R[G]$ is defined by the use of
the  multiplications in $G$ and $R$ as follows:

\[\left(\sum_{g_i \in G}a_ig_i\right)\left(\sum_{g_i \in G}b_ig_i\right) = \sum_{g_i\in G}\left(\sum_{g_jg_k = g_i}a_j b_k\right)g_i.\]

As an example of a group ring, we consider the symmetric group $S_5$
and the  ring $\mathbb{Z}_7$ and form the group ring
$\mathbb{Z}_7[S_5]$.
We will write the
identity element of $S_m$ as $e$. Sample elements and operations are
\begin{align*}
a&= 5(123)+2(15)(24)+(153) \\
b&= 3(123)+4(1453)\\
a+b &= (123)+2(15)(24)+(153)+4(1453) \\
ab &= (5(123)+2(15)(24)+(153))(3(123)+4(1453)) \\
 &= 15(132)+20(145)(23)+6(14235)+8(124)(35)+3(12)(35)+4(1435)\\
 &= (132)+6(145)(23)+6(14235)+(124)(35)+3(12)(35)+4(1435)\\
ba &= (3(123)+4(1453))(5(123)+2(15)(24)+(153)) \\
 &= 15(132)+6(15243)+3(15)(23)+20(12)(345)+8(13)(254)+4(1345) \\
 &= (132)+6(15243)+3(15)(23)+6(12)(345)+(13)(254)+4(1345)
\end{align*}

Now that group rings have been defined, it is clear how to define
$M_2(\mathbb{Z}_n[S_m])$,  the ring of $2 \times 2$ matrices over
the group ring $\mathbb{Z}_n[S_m]$. We are only going to be
concerned with multiplication of matrices in this ring; as an
example using the same $a$ and $b$ defined above, we can define
\[
M_1 =
\left[ {\begin{array}{cc}
 a & e  \\
 e & b  \\
 \end{array} } \right]
\text{, }M_2 =
\left[ {\begin{array}{cc}
 b & e  \\
 0 & a  \\
 \end{array} } \right].
\]
Then
\begin{align*}
M_1M_2 =&
\left[ {\begin{array}{cc}
 ab & 2a  \\
 b & e+ba  \\
 \end{array} } \right]\\
 =&
\left[ {\begin{array}{cc}
 ab & 3(123)+4(15)(24)+2(153) \\
 3(123)+4(1453) & e+ba \\
 \end{array} } \right],
 \end{align*}
where $ab$ and $ba$ are computed above.

\section{Computational Diffie-Hellman and Decision Diffie-Hellman}
\label{DDH}

Recall that in the Diffie-Hellman key exchange Alice and Bob want to
establish a  secret shared key. Alice chooses a finite group $G$ and
an element $g$  of the group $G$. Alice then picks a random $a$ and
publishes $(g,G,g^a)$. Bob also picks a random $b$ and publishes
$(g^b)$. Alice's and Bob's secret key is now $g^{ab}$, which can be
computed by both of them since $g^{ab} = (g^a)^b = (g^b)^a$. The
security of the Diffie-Hellman key exchange relies on the assumption
that it is computationally hard to recover $g^{ab}$ given $(g, G,
g^a, g^b)$.

A passive eavesdropper, Eve,  would try  to recover $g^{ab}$ from
$(g, G, g^a, g^b)$. One defines the Diffie-Hellman algorithm by
$F(g,G,g^a,g^b)=g^{ab}$. We say that a group $G$ satisfies the
Computational Diffie-Hellman (CDH) assumption if no efficient
algorithm exists to compute $F(g,G,g^a,g^b)=g^{ab}$. More precisely,

\begin{definition}

A CDH algorithm $F$ for a group $G$ is a probabilistic polynomial
time algorithm  satisfying, for some fixed $\alpha>0$ and all
sufficiently large $n$,
\[\mathbb{P}[F(g,G,g^a,g^b)=g^{ab}]>\frac{1}{n^{\alpha}}.\]
The probability is over a uniformly random choice of $a$ and $b$. We
say that the group $G$  satisfies the CDH assumption if there is no
CDH algorithm for $G$.

\end{definition}

Even though a group may satisfy the CDH assumption, CDH by itself is
not sufficient  to prove that the Diffie-Hellman protocol is useful
for practical cryptographic purposes. While Eve may not be able to
recover the entire secret, she may still be able to recover valuable
information about it. For example, even if CDH is true, Eve may
still be able to predict 80\% of the bits of $g^{ab}$ with
reasonable confidence \cite{Bon1}.

Hence if we are using $g^{ab}$ as the shared secret key, one must be
able to  bound the information Eve can extract about it given $g$,
$g^a$ and $g^b$. This is formally expressed by the much stronger
Decision Diffie-Hellman (DDH) assumption.

\begin{definition}

A DDH algorithm $F$ for a group $G$ is a probabilistic polynomial
time algorithm  satisfying, for some fixed $\alpha>0$ and all
sufficiently large $n,$
\[\left|\mathbb{P}[F(g,G,g^a,g^b,g^{ab})=``True"]-\mathbb{P}[F(g,G,g^a,g^b,g^c)=``True"]\right|>\frac{1}{n^{\alpha}}.\]
The probability is over a uniformly random choice of $a, b$ and $c$.
We say that the group $G$  satisfies the DDH assumption if there is
no DDH algorithm for $G$.

\end{definition}

Essentially, the DDH assumption implies that there is no efficient
algorithm which can distinguish between the two probability
distributions $(g^a,g^b,g^{ab})$ and $(g^a,g^b,g^c)$, where $a, b$
and $c$ are chosen at random.

\section{Diffie-Hellman key exchange protocol using matrices over $\mathbb{Z}_n[S_m]$}

While $S_m$ is a relatively small group for small $m$, the size of
the group ring $\mathbb{Z}_n[S_m]$   grows reasonably fast, even for
small values of $n$ and $m$. This is one reason we chose to look at
the Diffie-Hellman key exchange protocol using these group rings. We
propose to work with the group ring  $\mathbb{Z}_7[S_5]$, which has
the size $7^{5!}=7^{120}$. The next step is to work with matrices
over these group rings. Hence, say, the semigroup
$M_3(\mathbb{Z}_7[S_5])$ of $3\times 3$ matrices has the order
$(7^{5!})^9 \approx 10^{913}$. This semigroup of matrices can now
serve as the platform for the Diffie-Hellman key exchange protocol.
The procedure Alice and Bob carry out is essentially the same.

Alice chooses a public matrix $M \in M_3(\mathbb{Z}_7[S_5])$   and a
private large positive integer $a$,  computes $M^a$, and publishes
$(M, M^a)$. Bob chooses another large integer $b$, and computes and
publishes ($M^b$). Both Alice and Bob can now compute the same
shared secret key $K = (M^a)^b = (M^b)^a$.

As we have already mentioned in the Introduction, multiplication of
matrices in the semigroup $M_3(\mathbb{Z}_7[S_5])$ is very efficient,
and, of course, in this semigroup, as in any other semigroup, we can
use the ``square and multiply algorithm'' for exponentiation.


To assess security of our proposal,   we should address the two
Diffie-Hellman assumptions, CDH and DDH. We   investigate the
(stronger) DDH assumption experimentally in Section
\ref{experiments}.
%

Finally, some of the algebraic properties of
$M_3(\mathbb{Z}_7[S_5])$ will be investigated.

\section{Experimental results}
\label{experiments}

The CDH assumption  can only be answered theoretically, but the DDH
assumption  can be investigated experimentally. To construct our
matrix semigroups we implemented the necessary group ring procedures
in C++. We have the choice of which symmetric group to use and which
ring $\mathbb{Z}_n$ to use as well. Next we used a standard uniform
distribution implementation to allow for a random selection of an
element from our group ring. Finally, we constructed random $k\times
k$ matrices over our group ring. Experiments were carried out with
various group rings $M_k(\mathbb{Z}_n[S_m])$.

We propose the use of $S_5$ as the group for our experiments since
its underlying structure is understood and simple. When constructing
the semigroup $\mathbb{Z}_n[S_5]$,
 one has the benefits of using the group $S_5$ as a
building block. Namely, the group  $S_5$ has the advantage of having only
one normal subgroup, $A_5$, which has index 2 in $S_5$. Hence,
trying to get some information about $a$ from $M^a$ by applying a
non-trivial group homomorphism is limited only to the \textit{sign
homomorphism} $S_5 $ to $\mathbb{Z}_2$ of a symmetric group.

We naturally implemented a ``square and multiply" routine to speed
up computations for exponentiation.  With this procedure we can
compute high powers of random matrices from our matrix semigroups
fairly quickly, see Table \ref{speeds}.

We note that the computations were carried out on an Intel Core2 Duo
2.26GHz machine,  utilizing only one core, with 4GB of memory and
the times were computed as an average time after $250$ such
exponentiations. No optimizations were in effect and only one
processor was used. Thus computational time may be reduced
significantly by using more than one core and by implementing any
available optimizations for DH using our scheme.

As a comparison for computational times, we refer to recent results
of \cite{google} claiming new speed records for DH  implementations.
In the paper,  an implementation of the DH signature exchange
protocol  over the elliptic curve P-224 is presented. Without any
optimization they can carry out 1800 operations per second for the
DH protocol,  on a somewhat  more powerful computer than ours.
Recall that in P-224 you require approximately 340 operations for a
single ``exponentiation''. Hence, they require about 0.2 seconds per
DH exponentiation versus our 0.6 seconds in
$M_2(\mathbb{Z}_7[S_5])$.

    \begin{center}
    \small
    \begin{table}[h!]
    \caption{Speed of Computation}
    \begin{tabular}{c | c | c | r}
        Matrix Size & $\mathbb{Z}_n$ & Exponent & Avg. Time (s)\\
        \hline
        $2\times 2$ & 2 & $10^{10}$ & 0.06\\
        $2\times 2$ & 3  & $10^{10}$ & 0.06\\
        $2\times 2$ & 5  & $10^{10}$ & 0.06\\
        $2\times 2$ & 7  & $10^{10}$ & 0.06\\
        \hline

        $2\times 2$ & 2 & $10^{100}$ & 0.58\\
        $2\times 2$ & 3  & $10^{100}$ & 0.58\\
        $2\times 2$ & 5  & $10^{100}$ & 0.58\\
        $2\times 2$ & 7  & $10^{100}$ & 0.59\\
        \hline

        $2\times 2$ & 2 & $10^{1000}$ & 5.97\\
        $2\times 2$ & 3  & $10^{1000}$ & 6.11\\
        $2\times 2$ & 5  & $10^{1000}$ & 5.98\\
        $2\times 2$ & 7  & $10^{1000}$ & 6.66\\
        \hline

        $3\times 3$ & 2 & $10^{10}$ & 0.19\\
        $3\times 3$ & 3  & $10^{10}$ & 0.20\\
        $3\times 3$ & 5  & $10^{10}$ & 0.20\\
        $3\times 3$ & 7  & $10^{10}$ & 0.20\\
        \hline

        $3\times 3$ & 2 & $10^{100}$ & 1.95\\
        $3\times 3$ & 3  & $10^{100}$ & 1.95\\
        $3\times 3$ & 5  & $10^{100}$ & 1.94\\
        $3\times 3$ & 7  & $10^{100}$ & 1.94\\
        \hline

        $3\times 3$ & 2 & $10^{1000}$ & 20.17\\
        $3\times 3$ & 3  & $10^{1000}$ & 20.15\\
        $3\times 3$ & 5  & $10^{1000}$ & 19.72\\
        $3\times 3$ & 7  & $10^{1000}$ & 19.74\\
        \hline
    \end{tabular}

    \label{speeds}
    \end{table}
    \end{center}

One additional thing we noticed was that the speed of computation is
independent  of the number of nonzero terms in the  entries of our
matrices $M$. One possible intuitive explanation is based on the
fact that any symmetric group can be generated by a set of 2 particular
elements. Since we selected 9 (or 4) random group ring elements for each
 matrix, there is a high probability that we have
selected a pair of group elements that will generate all of our
symmetric group. Once we have multiplied $M$ by itself a few times
we get group ring elements of random length mixing throughout the
matrix entries.

Random group ring elements from $\mathbb{Z}_2[S_5]$ have coefficients
either 0 or 1  for each of the 120 elements of $S_5$. A simple
binomial distribution calculation shows that with probability around
$93\%$ a random element of this group ring has a total number of
nonzero terms between $50$ and $70$.

\subsection{Experimental results on the Decision Diffie-Hellman assumption}

We should note that for those experiments that were carried out
using $2\times 2$ matrices,  it is reasonable to assume that if the
results hold in the smaller matrix size, they will also hold for
$3\times 3$ matrices. In order to test the DDH assumption we need to
look at the two distributions: one generated by $(M^{a}, M^{b},
M^{ab})$ and the other generated by $(M^{a}, M^{b}, M^c)$ for a
random $c$. Ideally, we would like the two distributions to be
indistinguishable.

To verify that, we have run the following 3 experiments. In the
first experiment, we verify that,  as the common sense suggests,
$M^{ab}$ has the same distribution as $M^c$. In the second
experiment, we verify that $M^a$ is distributed ``uniformly", i.e.,
like a randomly selected matrix $N$. A ``randomly selected" matrix
here means a matrix whose entries are random elements of the
platform group ring. In turn, a random element of the group ring is
selected by selecting each coefficient uniformly randomly from the
ring of coefficients (in our case, from $\mathbb{Z}_7$).

Combining the results of these two experiments, we see that each
component in the triple $(M^{a}, M^{b}, M^{ab})$ is uniformly
distributed (for random $a, b$) in the sense described above. Now
our final experiment verifies that the whole triple $(M^{a}, M^{b},
M^{ab})$ is distributed like a triple of independently selected
random matrices $(N_1, N_2, N_3)$, and therefore the distribution is
indistinguishable from that of $(M^{a}, M^{b}, M^c)$ since the
latter, too, is distributed like a triple of independently selected
random matrices according to the previous experiments.

A more detailed description of the three experiments is below.

In the first experiment, we picked $a$ and $b$ randomly from the
interval $[10^{22},10^{28}]$, and $c$ randomly from
$[10^{44},10^{55}]$, so that $c$ had about the same size as the
product $ab$. To get a clearer picture of how different or similar
these final matrices were, we looked at each entry of the matrix.
For each choice of a random matrix $M$ and random $a, b,$ and $c$ we
computed the matrices $M^{ab}$ and $M^c$. This was repeated 500
times and we created a table that was updated after each run with
the distribution of elements of $S_5$ for each entry of the matrix.
We were working with $M_2(\mathbb{Z}_7[S_5])$.

After  $500$ runs we created Q-Q plots of entries of $M^{ab}$ versus
entries of $M^c$, where we use the  notation $M = \left(
\begin{smallmatrix} a_1&a_2 \\ a_3&a_4  \end{smallmatrix} \right)$. Q-Q plots
(or quantile plots) are a graphical method of comparing the
quantiles of the cumulative distribution function (cdf) $F$ versus
the corresponding quantiles of the cdf $G$. The functions are
parameterized by $p$, where $p \in [0,1]$. One axis represents
$F^{-1}(p)$ and the other axis represents $G^{-1}(p)$. If the two
cdf's are identical, then the Q-Q plot will be that of $y=x$. It
will also be a straight line if the distributions are of the same
type, but have different mean and standard deviation, see
\cite{Gibb} for more details.

\begin{figure}[ht]
\includegraphics[width=0.75\textwidth]{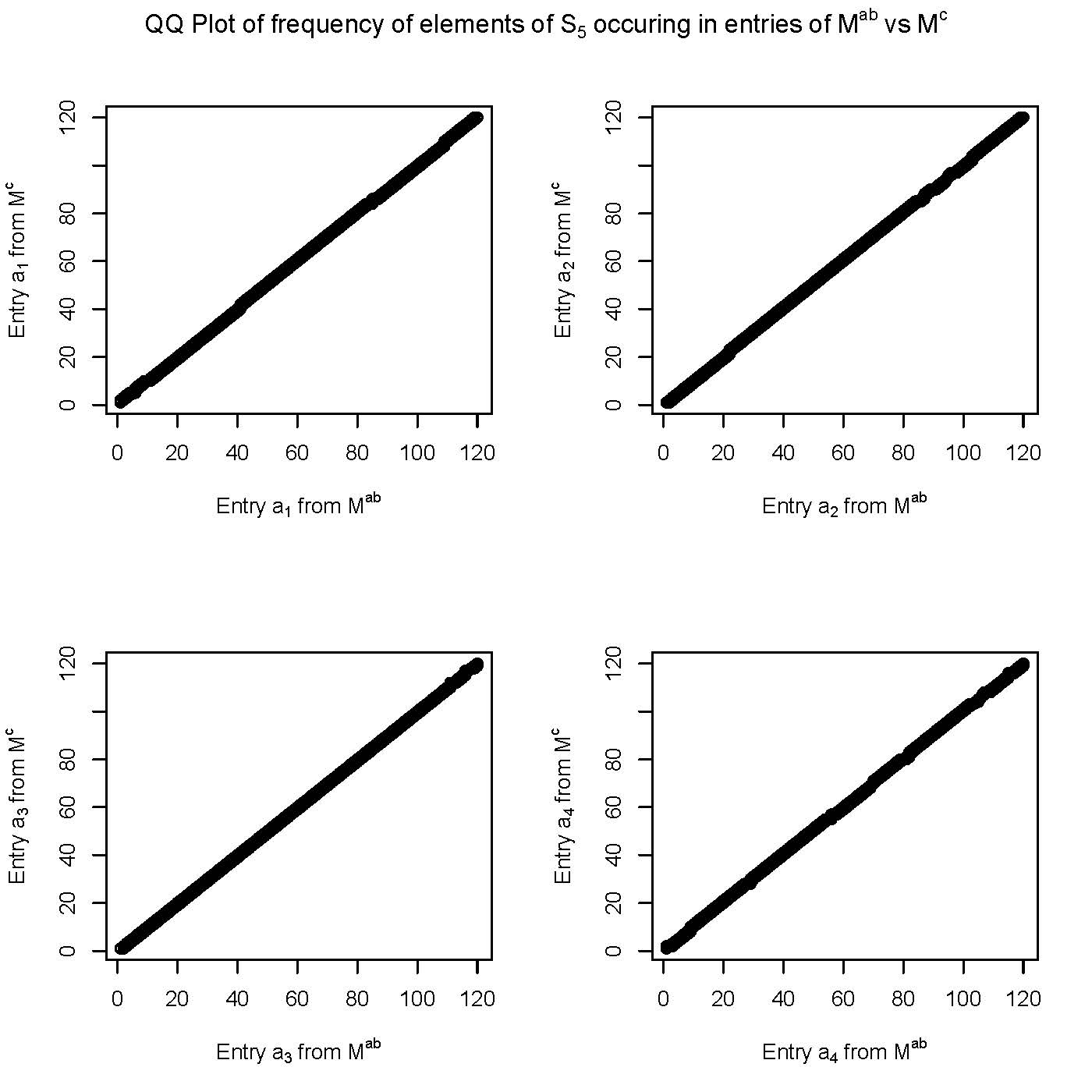}
\caption{DDH results for $M^{ab}$ vs. $M^c$}
\label{DDH_fig}
\end{figure}

As can be seen from Figure \ref{DDH_fig}, it appears that the
distributions of each of  the matrices $M^{ab}$ and $M^c$ are indeed
identical, which experimentally confirms what the common sense
suggests.

\begin{figure}[ht]
\includegraphics[width=0.75\textwidth]{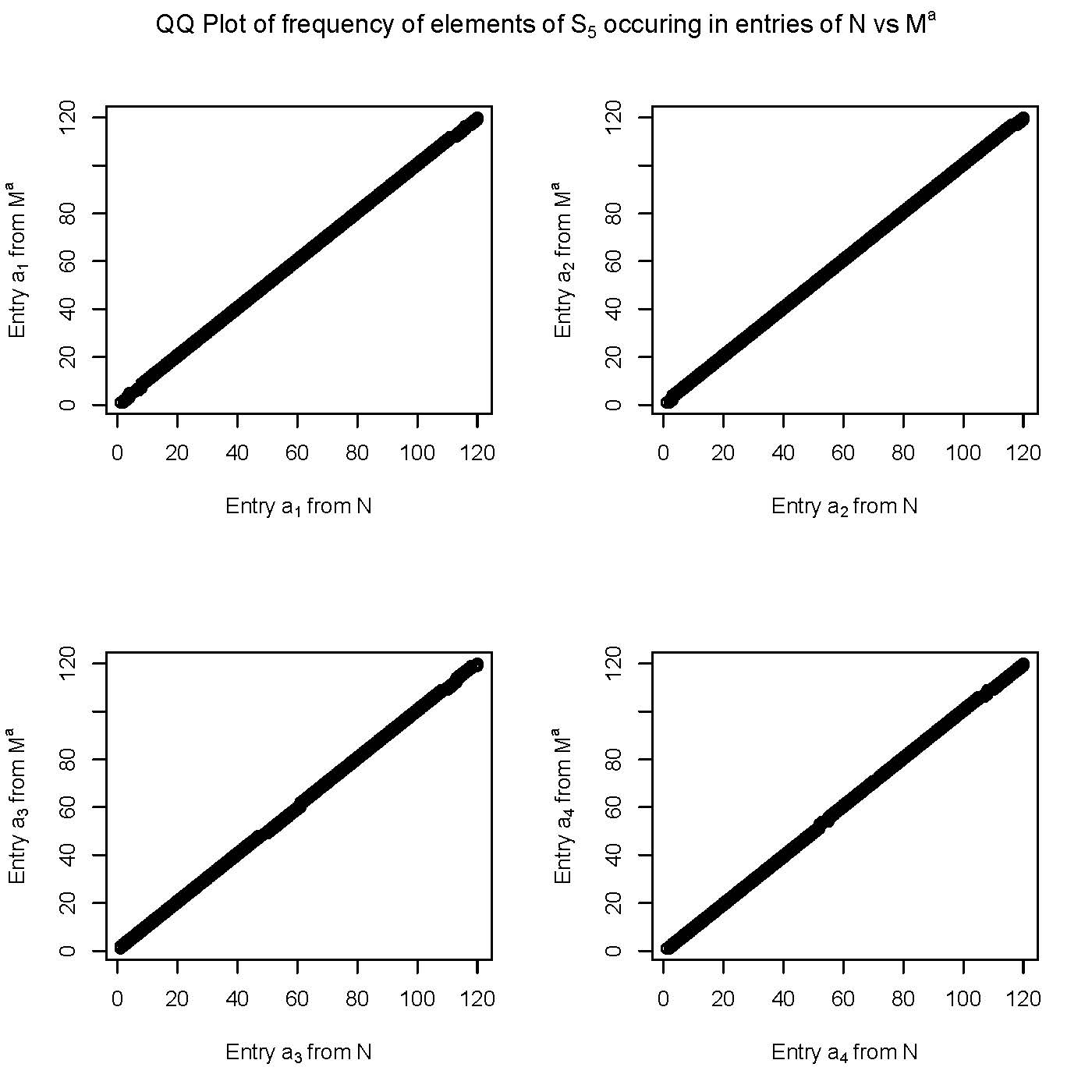}
\caption{DDH results for $N$ vs. $M^a$}
\label{DDH_fig2}
\end{figure}

In the second experiment, we verify that $M^a$ is distributed
``uniformly", i.e., like a randomly selected matrix $N$. We also
verify thereby that no information is leaked about $a$ by publishing
$M^a$, for a given $M$. The experimental setup was similar to the
previous one, only here we  chose two random matrices $M$ and $N$,
and a random integer $a \in$ $[10^{44},10^{55}]$. Again we produced
a Q-Q plot for  the two distributions, see Figure \ref{DDH_fig2}.
From the plot, it is clear that $M^a$ is indistinguishable from a
random matrix $N$.

Finally, we ran a third experiment to ensure the independence of
matrix entries from one another in the triple $(M^a, M^b, M^{ab})$
by comparing its distribution to that of the triple of independently
selected random matrices $(N_1, N_2, N_3)$. This is a valid and
important question to ask as the information contained within the
first two elements of the triple, which were shown to be random
previously, may  affect $M^{ab}$ in a predictable way. To this end,
we ran $30,000$ experiments four times, where for each element of
$S_5$ we counted the frequency of coefficients of $\mathbb{Z}_7$
that occurred in the entries of each of the matrices in $(M^a, M^b,
M^{ab})$. We used the same $M$ in each experiment, but varied $a$
and $b$.

More specifically, we formed triples (one entry for each entry of
the triple of matrices) consisting of the concatenation of the
coefficients in the respective entry of the matrices for the same
element of $S_5$. For example, if the coefficient at the same
element of $S_5$ in the upper left corner entry of the first matrix
is 0, in the second matrix it is 5, and in the third matrix it is 1,
then the concatenated coefficient is 051. Thus, there is a total of
$7^3=343$ concatenated coefficients.

We counted the occurrence of such triples throughout the experiments
for random choices of $a$ and $b$ in the same range as in the
previous experiments. We hypothesized that these coefficient triples
would be uniformly distributed over $\mathbb{Z}_7^3$, each occurring
with probability $1/7^3$. Since we performed $30,000$ such
experiments (four times), we anticipated that each element of this
distribution would show up approximately $30,000/7^3\sim 87$ times.

We reproduced a section of these results in the Table
\ref{distr_triples}, where we only used a portion of the table for
the $a_{11}$ entry of the matrices because of the space constraints.
Results for other entries are similar. The columns represent
elements of $S_5$ (i.e., in the full table there would be 120
columns), the rows represent concatenated coefficients of the
triples from $\mathbb{Z}_7^3$ (i.e., in the full table there would
be  $7^3=343$ rows), and the values in the table show the frequency
of occurrence of the coefficients. All tables have the same
``random'' structure, and it can be seen that there appears to be no
particular skew in the expected uniformity of the distribution of
these coefficients, which allows us to conclude that the
distribution of triples of all respective coefficients in $(M^a,
M^b, M^{ab})$ is, indeed, uniform on $\mathbb{Z}_7^3$. Since each
component in the triple is itself uniformity distributed (as
evidenced by our first two experiments), it follows that $M^{ab}$ is
distributed independently of $(M^a, M^b)$.

\begin{center}
\small
\begin{longtable}[htbp]{c|cccccccccccccccccc}
\caption{Distribution of coefficient triples}
\label{distr_triples}\\
& $s_{1}$ & $s_{2}$ & $s_{3}$ & $s_{4}$ & $s_{5}$ & $s_{6}$ & $s_{7}$ & $s_{8}$ & $s_{9}$ & $s_{10}$ & $s_{11}$ & $s_{12}$ & $s_{13}$ & $s_{14}$ & $s_{15}$ & $s_{16}$ & $s_{17}$ & $s_{18}$\\
\hline
\endfirsthead
& $s_{1}$ & $s_{2}$ & $s_{3}$ & $s_{4}$ & $s_{5}$ & $s_{6}$ & $s_{7}$ & $s_{8}$ & $s_{9}$ & $s_{10}$ & $s_{11}$ & $s_{12}$ & $s_{13}$ & $s_{14}$ & $s_{15}$ & $s_{16}$ & $s_{17}$ & $s_{18}$\\
\hline
\endhead
    $e_{1}$ & 87    & 90    & 83    & 90    & 86    & 93    & 85    & 84    & 88    & 88    & 93    & 88    & 85    & 77    & 88    & 94    & 93    & 91  \\
    $e_{2}$ & 79    & 78    & 90    & 89    & 92    & 74    & 87    & 88    & 87    & 86    & 95    & 93    & 84    & 88    & 92    & 89    & 90    & 87  \\
    $e_{3}$ & 86    & 86    & 83    & 89    & 95    & 91    & 93    & 90    & 94    & 85    & 82    & 87    & 84    & 84    & 86    & 84    & 89    & 89  \\
    $e_{4}$ & 94    & 83    & 87    & 91    & 86    & 91    & 86    & 84    & 89    & 94    & 87    & 88    & 87    & 89    & 90    & 89    & 88    & 84  \\
    $e_{5}$ & 81    & 88    & 82    & 85    & 85    & 94    & 86    & 89    & 92    & 84    & 94    & 90    & 93    & 86    & 83    & 79    & 93    & 85  \\
    $e_{6}$ & 87    & 81    & 92    & 84    & 85    & 89    & 93    & 83    & 79    & 80    & 95    & 95    & 86    & 83    & 93    & 89    & 88    & 87  \\
    $e_{7}$ & 87    & 84    & 82    & 91    & 96    & 88    & 88    & 81    & 97    & 89    & 88    & 86    & 90    & 90    & 93    & 85    & 96    & 88  \\
    $e_{8}$ & 79    & 89    & 89    & 83    & 92    & 87    & 88    & 83    & 92    & 91    & 82    & 90    & 86    & 88    & 89    & 89    & 91    & 87  \\
    $e_{9}$ & 82    & 79    & 83    & 86    & 88    & 81    & 90    & 93    & 88    & 89    & 87    & 85    & 88    & 91    & 85    & 90    & 87    & 92  \\
    $e_{10}$ & 79    & 90    & 85    & 81    & 84    & 84    & 84    & 91    & 87    & 90    & 75    & 88    & 95    & 90    & 80    & 87    & 90    & 90  \\
    $e_{11}$ & 90    & 80    & 96    & 90    & 78    & 89    & 86    & 87    & 91    & 83    & 90    & 88    & 93    & 94    & 92    & 85    & 80    & 90  \\
    $e_{12}$ & 89    & 91    & 93    & 86    & 86    & 90    & 93    & 94    & 91    & 94    & 87    & 87    & 89    & 85    & 85    & 87    & 82    & 79  \\
    $e_{13}$ & 90    & 81    & 90    & 87    & 88    & 89    & 89    & 83    & 85    & 87    & 86    & 92    & 93    & 87    & 94    & 81    & 94    & 90  \\
    $e_{14}$ & 84    & 88    & 89    & 86    & 89    & 98    & 90    & 89    & 88    & 81    & 88    & 85    & 84    & 87    & 82    & 91    & 89    & 90  \\
    $e_{15}$ & 86    & 86    & 86    & 87    & 94    & 95    & 90    & 88    & 85    & 84    & 86    & 83    & 87    & 90    & 92    & 92    & 88    & 88  \\
    $e_{16}$ & 86    & 87    & 80    & 81    & 81    & 95    & 88    & 86    & 84    & 88    & 91    & 95    & 92    & 82    & 86    & 89    & 87    & 83  \\
    $e_{17}$ & 80    & 87    & 86    & 87    & 91    & 80    & 94    & 87    & 86    & 97    & 82    & 85    & 85    & 91    & 91    & 89    & 93    & 89  \\
    $e_{18}$ & 84    & 89    & 82    & 89    & 91    & 89    & 88    & 92    & 81    & 82    & 92    & 88    & 82    & 87    & 88    & 84    & 87    & 81  \\
    $e_{19}$ & 78    & 88    & 85    & 83    & 92    & 84    & 86    & 97    & 86    & 89    & 87    & 87    & 80    & 87    & 92    & 87    & 94    & 88  \\
    $e_{20}$ & 91    & 95    & 85    & 89    & 94    & 86    & 96    & 88    & 88    & 91    & 82    & 89    & 78    & 90    & 88    & 89    & 89    & 87  \\
    $e_{21}$ & 85    & 89    & 87    & 82    & 88    & 85    & 89    & 94    & 79    & 81    & 86    & 86    & 80    & 86    & 89    & 86    & 90    & 81  \\
    $e_{22}$ & 85    & 92    & 86    & 83    & 87    & 85    & 84    & 78    & 81    & 85    & 83    & 89    & 92    & 95    & 93    & 90    & 90    & 87  \\
    $e_{23}$ & 84    & 91    & 86    & 86    & 83    & 88    & 84    & 89    & 88    & 82    & 95    & 90    & 87    & 90    & 84    & 79    & 82    & 81  \\
    $e_{24}$ & 97    & 83    & 93    & 93    & 90    & 91    & 88    & 95    & 86    & 87    & 88    & 94    & 83    & 88    & 86    & 99    & 94    & 85  \\
    $e_{25}$ & 88    & 83    & 92    & 88    & 85    & 82    & 90    & 82    & 88    & 86    & 92    & 87    & 86    & 86    & 87    & 83    & 84    & 88  \\
    $e_{26}$ & 86    & 89    & 78    & 85    & 93    & 87    & 85    & 85    & 84    & 87    & 87    & 94    & 102   & 86    & 93    & 91    & 91    & 92  \\
    $e_{27}$ & 90    & 83    & 77    & 81    & 94    & 85    & 86    & 83    & 90    & 86    & 87    & 92    & 90    & 82    & 79    & 95    & 83    & 85  \\
    $e_{28}$ & 85    & 79    & 86    & 83    & 80    & 85    & 88    & 88    & 85    & 86    & 92    & 94    & 88    & 87    & 84    & 92    & 84    & 91  \\
    $e_{29}$ & 91    & 94    & 86    & 92    & 88    & 82    & 93    & 85    & 88    & 93    & 88    & 92    & 85    & 92    & 77    & 87    & 89    & 88  \\
    $e_{30}$ & 97    & 91    & 88    & 87    & 88    & 88    & 81    & 87    & 89    & 89    & 82    & 81    & 82    & 94    & 84    & 87    & 87    & 91  \\
    $e_{31}$ & 89    & 91    & 92    & 87    & 97    & 88    & 89    & 83    & 89    & 92    & 84    & 84    & 78    & 89    & 81    & 101   & 83    & 86  \\
    $e_{32}$ & 88    & 84    & 81    & 90    & 80    & 91    & 90    & 89    & 89    & 87    & 89    & 83    & 93    & 91    & 100   & 87    & 88    & 87  \\
    $e_{33}$ & 82    & 90    & 81    & 86    & 94    & 93    & 93    & 91    & 88    & 88    & 85    & 85    & 79    & 92    & 82    & 87    & 84    & 87  \\
    $e_{34}$ & 87    & 91    & 91    & 92    & 86    & 85    & 94    & 85    & 79    & 94    & 82    & 80    & 87    & 89    & 89    & 86    & 93    & 90  \\
    $e_{35}$ & 85    & 90    & 88    & 83    & 88    & 82    & 90    & 92    & 88    & 88    & 90    & 91    & 77    & 90    & 91    & 90    & 87    & 91  \\
    $e_{36}$ & 79    & 90    & 89    & 86    & 95    & 90    & 89    & 87    & 90    & 84    & 93    & 91    & 85    & 84    & 80    & 94    & 93    & 84  \\
    $e_{37}$ & 90    & 81    & 82    & 97    & 87    & 92    & 89    & 81    & 80    & 88    & 91    & 92    & 94    & 90    & 86    & 81    & 83    & 96  \\
    $e_{38}$ & 95    & 91    & 91    & 86    & 79    & 91    & 93    & 83    & 82    & 87    & 86    & 92    & 89    & 83    & 94    & 92    & 85    & 85  \\
    $e_{39}$ & 86    & 87    & 90    & 84    & 96    & 80    & 89    & 82    & 90    & 86    & 91    & 84    & 80    & 79    & 82    & 96    & 98    & 91  \\
    $e_{40}$ & 92    & 87    & 92    & 80    & 84    & 91    & 90    & 88    & 91    & 92    & 86    & 81    & 86    & 92    & 86    & 90    & 92    & 87  \\
    $e_{41}$ & 84    & 90    & 90    & 91    & 83    & 86    & 91    & 90    & 88    & 84    & 88    & 86    & 89    & 82    & 83    & 92    & 92    & 90  \\
    $e_{42}$ & 90    & 90    & 86    & 76    & 96    & 86    & 87    & 80    & 89    & 83    & 87    & 99    & 88    & 89    & 84    & 90    & 89    & 86  \\
    $e_{43}$ & 81    & 86    & 97    & 83    & 89    & 84    & 88    & 88    & 83    & 84    & 96    & 87    & 87    & 90    & 91    & 82    & 91    & 87  \\
    $e_{44}$ & 86    & 82    & 90    & 89    & 76    & 87    & 93    & 81    & 83    & 91    & 85    & 88    & 90    & 86    & 90    & 90    & 84    & 90  \\
    $e_{45}$ & 88    & 95    & 88    & 88    & 95    & 91    & 83    & 92    & 92    & 86    & 82    & 82    & 94    & 87    & 88    & 92    & 83    & 90  \\
    $e_{46}$ & 93    & 87    & 96    & 80    & 89    & 90    & 86    & 84    & 87    & 100   & 85    & 95    & 89    & 93    & 96    & 84    & 91    & 85  \\
    $e_{47}$ & 92    & 85    & 85    & 85    & 91    & 91    & 87    & 88    & 83    & 89    & 87    & 85    & 89    & 83    & 89    & 86    & 84    & 83  \\
    $e_{48}$ & 90    & 87    & 82    & 99    & 76    & 82    & 84    & 82    & 83    & 95    & 83    & 94    & 92    & 87    & 93    & 86    & 86    & 82  \\
    $e_{49}$ & 93    & 82    & 85    & 86    & 85    & 87    & 91    & 85    & 80    & 91    & 94    & 87    & 92    & 90    & 90    & 87    & 86    & 96  \\
    $e_{50}$ & 90    & 78    & 85    & 83    & 85    & 88    & 93    & 82    & 84    & 87    & 92    & 82    & 84    & 89    & 85    & 81    & 84    & 88  \\
    $e_{51}$ & 90    & 90    & 83    & 86    & 97    & 87    & 88    & 90    & 90    & 92    & 88    & 85    & 96    & 86    & 90    & 90    & 88    & 98  \\
    $e_{52}$ & 88    & 82    & 92    & 92    & 88    & 83    & 94    & 92    & 91    & 92    & 89    & 89    & 87    & 91    & 81    & 81    & 87    & 88  \\
    $e_{53}$ & 88    & 89    & 89    & 86    & 92    & 86    & 85    & 86    & 90    & 93    & 75    & 90    & 91    & 95    & 87    & 84    & 92    & 83  \\
    $e_{54}$ & 89    & 91    & 88    & 92    & 82    & 84    & 95    & 84    & 82    & 82    & 85    & 86    & 91    & 93    & 93    & 96    & 84    & 75  \\
    $e_{55}$ & 88    & 88    & 84    & 88    & 82    & 94    & 100   & 89    & 84    & 88    & 79    & 89    & 90    & 85    & 88    & 83    & 85    & 85  \\
\end{longtable}
\end{center}

\subsection{Experimental results on low orbits}
\label{orbits}

Here we address the following ``low orbits" question: we want to
make sure that powers of the public matrix $M$ in our semigroup do
not end up in an orbit of low order. This means that if Alice
chooses  a random integer $a$, we cannot have $M^n= M^k$, for $n < k
<<a$ (similarly for $b$ chosen by Bob). If this were the case, then
the eavesdropper Eve could first determine $n$ and $k$, then she
could find  the values of $c$ and $d$, where $1\leq c,d \leq k$,
such that $M^a=M^c$ and $M^b=M^d$. The shared secret key then could
be computed as
\[M^{ab}=(M^a)^b=(M^c)^b=(M^b)^c=(M^d)^c=M^{cd}.\] This is similar
to the problem of finding a generator (i.e., an element of maximum
order) in the multiplicative group of $\mathbb{Z}_p$, the original
platform for the Diffie-Hellman  protocol. Since we are dealing with
a semigroup (of matrices)  where most elements are not invertible
and therefore do not have an ``order" in the usual sense, we
consider those orbits instead.

While it is conceivable that for a random matrix from
$\mathbb{Z}_7[S_5]$ the length of such an orbit is going to be huge,
we realize that when we are providing Alice and Bob with a matrix
$M$, we have to at least have some solid lower bound for the length
of an orbit for powers of $M$. Here is one possible approach.

The matrix $M$ will be a product of two matrices:  $M = M_1 \cdot
S$, where $M_1$ is  a random invertible matrix from
$\mathbb{Z}_7[S_5]$, and $S$ is a ``scalar" matrix that has zeros
off the diagonal and each element on the diagonal is $s =
(3+g_1)(3+g_2)(3+g_3)(3+g_4)(3+g_5) (3+g_6)(5+h)$. Here $g_i$ are
elements of $S_5$ that generate different subgroups of order 5, and
$h$ is a product of a 2-cycle and a 3-cycle. The element $s$ is not
invertible because it is a zero divisor. To see this, write $(5+h)$
as $(h-2)$ and multiply it by $\sum_{i+j=5} h^i 2^j$ to get
$(h^6-2^6) = 0$ since  $h^6=2^6 = 1$ in our group ring. Therefore,
the matrix $S$ is not invertible either. We have run a computer
program trying to detect an orbit generated by powers of $S$. While
our program has not terminated in the allotted time (several weeks),
we know that there are no orbits up to $s^{10^{10}}$. Then, for a
random invertible matrix $M_1$, we have just computed powers of
$M_1$ up to $M_1^{10^{10}}$, and none of these powers was the
identity matrix (or even a diagonal matrix). We note that looking
for orbits going through powers of a {\it non-invertible} matrix $M$
would consume much more resources and was, in fact, infeasible
beyond $M^{10^6}$ given our computational resources. This is because
once each power of $M$ is computed, it needs to be stored and
eventually compared to all other powers of $M$. For an invertible
matrix $M_1$, on the other hand, we do not need to store any powers
to find its order.

Now we claim that with overwhelming probability, if we have a random
invertible matrix $M_1$  with the property that the powers of $M_1$
up to $10^{10}$ are not diagonal matrices, then the powers of $M_1
\cdot S$ up to $10^{10}$ do not have any orbits. To see this, let us
assume that the matrices $M_1$ and $S$ commute; if our claim is
valid under this assumption then it is also valid without  this
assumption since adding a relation $M_1S=SM_1$ is like considering a
homomorphic image: equalities will be preserved.

Suppose now that we have $(M_1S)^n = (M_1S)^{n+k}$ for some positive
integers $n, k$, with $k < 10^{10}$. If $M_1$ and $S$ commute, this
yields $M_1^n S^n = M_1^{n+k} S^{n+k}$. Since $M_1$ is invertible,
we can  cancel $M_1^n$ and get $S^n = M_1^{k} S^{n+k}$, and then

$$(M_1^{k} S^{k} -I) \cdot S^n = O,$$

\noindent where $I$ is the identity matrix and $O$ is the zero
matrix. While it is possible that the product of two nonzero
matrices is the zero matrix, the probability of this to happen is
negligible, given that the matrix $M_1^{k} S^{k} -I$ is not even
diagonal (with overwhelming probability) if $k < 10^{10}$, as our
experiments suggest. The matrix $S^n$, on the other hand, is
diagonal; therefore, for the displayed equality above to hold, every
non-zero element $a_{ij}$ of the matrix $(M_1^{k} S^{k} -I)$ has to
be a zero divisor such that $a_{ij} \cdot r =0,$ where $r$ is the
element on the diagonal of the matrix $S^n$ (the latter is obviously
a scalar matrix). This (somewhat informal) argument shows that $k >
10^{10}$ with overwhelming probability. We realize that this lower
bound may not be very impressive, but more convincing lower bounds
may be based on less convincing arguments. We believe that, in fact,
$k > 10^{80}$ with overwhelming probability, but at the time of this
writing we do not have a convincing argument to support that belief.

To conclude this section, we say a few words about sampling
invertible matrices. There are several techniques for doing this;
here we give a brief exposition of one of them. We start with an
already ``somewhat random" matrix, for which it is easy to compute
the inverse. An example of such a matrix is a lower/upper triangular
matrix, with invertible elements on the diagonal:

\vskip -0.5cm

\begin{align*}
U&=\begin{pmatrix}
g_1 & u_1 & u_2 \\
0 & g_2 & u_3 \\
0 & 0 & g_3 \\
\end{pmatrix}.
\end{align*}

Here $g_i$ are random elements of the group $S_5$, and $u_i$ are
random elements of the group ring $\mathbb{Z}_{7}[S_5]$. We then
take a random product, with 20 factors, of such random  invertible
upper and lower triangular matrices, to get our invertible  matrix
$M_1$.

\section {``Standard"  Attacks}
\label{attacks}

In this section, we discuss why three ``standard" attacks on the
``classical" discrete logarithm problem do not work with our
platform semigroup.

\subsection{Baby--step giant--step algorithm}
One known method of attacking the ``classical" discrete logarithm
problem, due to Shanks \cite{Shanks}, is the baby-step giant-step
algorithm. The algorithm computes discrete logarithms in a group of
order $q$ in $O\left(\sqrt{q} ~\text{polylog}(q)\right)$ time, where
$\text{polylog}(q)$ is $O((\log(q))^c)$ for some constant $c$. If
adapted to our situation, this
algorithm would look as follows.\\

\begin{center}
    \small
\begin{tabular}{|l|}
    \hline
    \textbf{Baby-step giant-step algorithm}\\\\
    \textbf{Input:} $M,$ $A\in M_3(\mathbb{Z}_7[S_5])$, $n=|M_3(\mathbb{Z}_7[S_5])|$\\
    \textbf{Output:} $x \in \mathbb{N}, \ni M^x = A$\\\\
    Set $s:=\lceil \sqrt{n} \rceil$\\
    Set $t:=\lceil n/s \rceil$ \\
    \textbf{for} $i=0$ to $s$\\
    \hspace{1cm} \textbf{compute and store} $(i,AM^i)$\\
    \textbf{for} $j=0$ to $t$\\
    \hspace{1cm} \textbf{compute} $M_j=M^{js}$\\
    \hspace{1cm} \textbf{if} $M_j=AM^{i}$, for some $i$, $\textbf{return}$ $js-i$\\ \\
    \hline
\end{tabular}
\end{center}

There are a couple of points that have to be made about this
algorithm. The first is that we  need to produce a good method of
storing the matrices. This could be possible with a hash function,
in which case insertion and lookup is constant in time. However, our
matrices are fairly complex objects, and we need to take into
account the storage requirements of the algorithm.

Furthermore, we should note that  the order of our chosen random
matrix $M$ is much smaller than that of the whole group ring. Hence,
it may be possible to use a smaller value of $n$ as an input.
However, this requires knowledge of the order of $M$. As little is
known about the structure of this group ring,  we are not guaranteed
that the order exists in the usual sense. We are basically back to
looking for  orbit collisions as in our Section \ref{orbits}.


Each  entry in the matrix can be represented by a sequence of 120
(three-bit) coefficients. We can use a 360 bit string where we
encode each three-bit sequence with the value of the coefficient of
that polynomial term in $\mathbb{Z}_7[S_5]$. Hence each matrix will
need $360\times 4$ bits of  storage. In this algorithm we are
required to store
$\sqrt{|M_3(\mathbb{Z}_7[S_5])|}=\sqrt{7^{540}}\sim 10^{456}$ such
matrices. In order to store all these matrices we would need
$1440\times 10^{456}$ bits of space. This works out to about
$10^{446}TB$ of (memory or hard drive) space. Thus, it looks like
this algorithm is infeasible already in terms of space. Of course,
storing the arrays can   be optimized, e.g. we do not need to store
entries with zeroes. However, the amount of information that we need
to store, $10^{456}$ matrices, is still too big even if we only
store the number of non-zero terms in the polynomials.

One approach often suggested to decrease space requirements is to
decrease $s$,  hence increasing $t$. In this case the algorithm
instead of running in $O(\sqrt{n})$ time will run in $O(n/t)$ time.
Every time we reduce by half the storage requirements, we end up
doubling the running time of the algorithm. However, regardless of
what $s$ and $t$ are chosen to be we still need to perform $s+t$
group operations in the two loops. Given our constraints, the number
of group operations is minimized when $s=t=\sqrt{n}$. Hence, we need
at least $10^{457}$ group operations to run this algorithm, which is
again computationally infeasible.

\subsection{Other attacks}

There are two other algorithms that have been suggested for solving
the ``classical" discrete logarithm problem. The first is the
Pohlig-Hellman algorithm \cite{PohHel}. This algorithm relies on the
order of a group element and the generalized Chinese remainder
theorem to break the problem into smaller subproblems.

Specifically, suppose the order of the element $g \in G$ is $q$. In
the Diffie-Hellman scheme we wish to find  an $x$ such that $g^x=y$.
Suppose we know a factorization \[q=\prod_{i=1}^n{q_i},\] where the
$q_i$ are relatively prime. Then we have
\[ \left(g^{q/q_i}\right)^x = \left(g^x\right)^{q/q_i}=y^{q/q_i}, \text{ for } i=1,...,n. \]

By the Chinese remainder theorem we can write
\[\mathbb{Z}_q \cong \mathbb{Z}_{q_1} \times \cdots \times \mathbb{Z}_{q_n} \]
and we are left to solve $n$ instances of the discrete logarithm
problem in the smaller groups, i.e., defining $g_i=g^{q/q_i}$, we
must find the solutions $\{x_i\}_{i=1}^n$ for which
$g_i^{x_i}=y^{q/q_i}=g^x$.

However, in our situation the order of matrices in
$M_3(\mathbb{Z}_7[S_5])$ does not relate to the size of the whole
ring $M_3(\mathbb{Z}_7[S_5])$. Again, under multiplication this ring
is a semigroup, not a group, and the proportion of invertible
elements in this semigroup is very small. Additionally, the size of
this ring is $7^{1080}$, so the Chinese remainder theorem does not
really help in breaking this problem into smaller parts. If,
however, there was a way to break the problem into smaller
subproblems, we would still need to solve the discrete logarithm
problem in our setting, which so far as we know can only be done via
brute force.

The second algorithm proposed for solving the ``classical" discrete
logarithm problem  is Pollard's rho algorithm \cite{Poll}. The
inputs are  group elements $M$ and $N$, and the output is an integer
$n$ such that $M^n=N$. The algorithm first looks for an orbit, which
has the general form $M^{a}N^{b}=M^{c}N^{d}$, for $a,b,c$ and $d \in
\mathbb{N}$. This is done by using Floyd's cycle-finding algorithm.
As long as $b \ne d$, one can take the logarithm with base $M$ to
determine $n$:
\begin{align*}
M^{a}N^{b}&=M^{c}N^{d} \\
\Rightarrow a+b\log_M N & = c + d\log_M N \\
\Rightarrow \frac{a-c}{d-b} & = \log_M N \\
\Rightarrow M^{\frac{a-c}{d-b}}&=N
\end{align*}


However, in applying Floyd's cycle-finding algorithm in Pollard's
rho attack, the knowledge of the order of the cyclic group generated
by $M$ is essential. In our situation, not only is the order of $M$
unknown, but more importantly, since a random $M$ is not going to be
invertible with overwhelming probability, order considerations are
not applicable, and therefore neither is Pollard's rho attack, at
least in its standard form.

\subsection{Quantum Algorithm Attacks}
\label{quant}

It is well known that many cryptographic protocols are vulnerable to
quantum algorithm attacks \cite{Shor}. In particular, the
Diffie-Hellman protocol can be attacked using Shor's algorithm. This
algorithm basically recasts the discrete logarithm problem as a
hidden subgroup problem (HSP) and uses the quantum algorithms
developed for HSP to recover the exponent.

We believe that our protocal is secure against such attacks. The HSP
relies on the existence  of a function $f:G\rightarrow S$, for some
set $S$, such that $f$ is constanct on cosets of the unknown
subgroup $H \le G$ and also takes on distinct values for each coset.
For the discrete log we define $f:\mathbb{Z}_N \times
\mathbb{Z}_N\rightarrow G$, such that $f(a,b) = g^ax^b$, where $a,b
\in \mathbb{Z}_N$, $g,x \in G$, $g^\alpha=x$ and $|g|=N$. We can
rewrite this as $f(a,b)=g^{a+ b \cdot log_g x}$, and hence $f$ is
constant on the sets $L_c =\{(a,b) | a + b\log_g x = c\}$.

In this setup the  hidden subgroup we are seeking is
\[H=L_0=\{(0,0),(\log_g x, -1),(2\log_g x,-2), \cdots, (N\log_g
x,-N)\}.\] To be able to apply this algorithm one would need to know
the order of a matrix. However, this is not known a priori and it is
also the case that invertible matrices are sparse in our setup.
Hence in our setup the function $f$ is ill-defined.

Furthermore, given a random non-invertible matrix it is unlikely
that the function  $f$ will be distinct on cosets of the subgroup
$H$ or even constant on the different cosets. To see this assume $M$
is a non-invertible matrix, then powers of $M$ will either end up in
an orbit or will eventually become the zero matrix.  If we are in an
orbit, assume for example that $M^9 = M^{15}$ and the exponent we
are seeking is $\alpha=12$. The subgroup we are trying to identify
is $H=\{(0,0),(12,-1),(24,-2),(36,-3),\cdots\}$. From the setup we
note that $(36,-3) \sim (18,-3)$, but $(18,-3) \notin H$, for if it
were then $(36,-3)-(18,-3)=(18,0)\in H$, which is a contradiction.
On the other hand, assume some power of $M$ is the zero matrix, say
$M^{20}=0$, and again $\alpha=12$. In this case $f$ is no longer
constant on the subgroup $H$ as $0=f(24,-2) \ne f(12,-1)=I$.

\section{Conclusions}

Our contribution here is proposing the semigroup of matrices (of a
small size, $2\times 2$ or $3\times 3$) over the  group ring
$\mathbb{Z}_7[S_5]$, with the usual matrix multiplication operation,
as the platform for the Diffie-Hellman key exchange scheme. What we
believe is the main advantage of our  platform  over the standard
$\mathbb{Z}_p^\ast$  platform in the original  Diffie-Hellman scheme
is that the multiplication of matrices over $\mathbb{Z}_7[S_5]$ is
very efficient. In particular, in our setup multiplying elements is
faster than multiplying numbers in $\mathbb{Z}_p$ for a large $p$.
This is due to the fact that one can pre-compute the multiplication
table for the group $S_5$ (of order 120), so  in order to multiply
two elements of $\mathbb{Z}_7[S_5]$ there is no ``actual"
multiplication involved, but just re-arrangement of a bit string of
length $3\times 120$.  Also, no reduction modulo a large $p$ is
involved.

To verify the security of using such a semigroup of matrices as the
platform, we have experimentally addressed the \textit{Decision
Diffie-Hellman} assumption (Section  \ref{experiments}) and showed,
by using Q-Q plots (or quantile plots) that after 500 runs of the
experiment, two distributions, one generated by $M^{ab}$ and the
other generated by $M^c$ for a random $c$, are indistinguishable,
thereby experimentally confirming the DDH assumption for our
platform. Furthermore, no information is leaked from $M^a$ by
comparing it to a random matrix $N$.


From the security point of view, the advantages of our platform over
$\mathbb{Z}_p$ also include the fact  that neither ``standard"
attacks (baby--step giant--step, Pohlig-Hellman, Pollard's rho) nor
quantum algorithm attacks work with our platform, as we showed in
Section \ref{attacks}.

\section{Appendix: a challenge}
\label{challenge}

Here we present a challenge relevant to our Diffie-Hellman-like
scheme: given explicit $3\times 3$ matrices $M$, $M^{a}$, and
$M^{b}$ over the group ring $\mathbb{Z}_2[S_5]$, recover the matrix $M^{ab}$. Note that our recommended platform ring
is actually $\mathbb{Z}_7[S_5]$, but we believe that breaking our challenge is currently infeasible even
for $\mathbb{Z}_2[S_5]$. \\

\begin{landscape}

Below are the entries for $M$:\\

\small
\noindent
$a_{11} =   \epsilon+(2 4 3)+(2 4)+(1 2 3 4)+(1 2 4 3)+(1 2 4)+(1 3)(2 4)+(1 3 2 4)+(1 4 3 2)+(1 4 2)+(1 4)+(2 3)+(1 3)+(3 5 4)+(2 3 5 4)+(2 4)(3 5)+(2 5 3)+(2 5 4)+(2 5)(3 4)+(1 2)(4 5)+(1 2)(3 4 5)+(1 2)(3 5 4)+(1 2)(3 5)+(1 2 3 4 5)+(1 2 3 5)+(1 2 4 5 3)+(1 2 4 5)+(1 2 4)(3 5)+(1 2 5 3 4)+(1 3 5 4 2)+(1 3 5 2)+(1 3 4 5)+(1 3)(2 5 4)+(1 3)(2 5)+(1 3 4)(2 5)+(1 3 4 2 5)+(1 4 5 2)+(1 4 2)(3 5)+(1 4 5 3)+(1 4 5)+(1 4 3 5)+(1 4)(2 3 5)+(1 4 2 3 5)+(1 4 2 5 3)+(1 4 3)(2 5)+(1 4)(2 5 3)+(1 4 3 2 5)+(1 4)(2 5)+(1 5 4 3 2)+(1 5 3 2)+(1 5 2)(3 4)+(1 5 4 2 3)+(1 5 4)(2 3)+(1 5)(2 3)+(1 5 2 3 4)+(1 5)(2 3 4)+(1 5 3)(2 4)+(1 5 3 2 4)+(1 5)(2 4 3)+(1 5)(2 4)$\\
\\$a_{21} = \epsilon+(2 4 3)+(2 4)+(1 2 4 3)+(1 3 4 2)+(1 3)(2 4)+(1 3 2 4)+(1 4 2)+(1 4 2 3)+(1 4)(2 3)+(1 3)+(4 5)+(2 3 4 5)+(2 4 5 3)+(2 4 5)+(2 5 3)+(2 5)+(2 5 3 4)+(2 5)(3 4)+(1 2)(3 4 5)+(1 2)(3 5)+(1 2 3)(4 5)+(1 2 3 5)+(1 2 4 5 3)+(1 2 4 5)+(1 2 5)+(1 2 5 3 4)+(1 3 4 5 2)+(1 3 5 4 2)+(1 3 5 2)+(1 3)(4 5)+(1 3 4 5)+(1 3 5)(2 4)+(1 3)(2 5 4)+(1 3 4)(2 5)+(1 4 5 2)+(1 4 2)(3 5)+(1 4 5 3)+(1 4 5)+(1 4)(3 5)+(1 4 3 5)+(1 4 5)(2 3)+(1 4 3)(2 5)+(1 4)(2 5)+(1 4 2 5)+(1 5 4 3 2)+(1 5 4 2)+(1 5 2)+(1 5 3)+(1 5 4)+(1 5 3 4)+(1 5)(3 4)+(1 5 4 2 3)+(1 5 2 3)+(1 5 4)(2 3)+(1 5)(2 3 4)+(1 5 3)(2 4)+(1 5 2 4 3)+(1 5 3 2 4)+(1 5)(2 4 3)+(1 5 2 4)$\\
\\$a_{31} = (2 4 3)+(1 2 4)+(1 3 4 2)+(1 3)(2 4)+(1 3 2 4)+(1 4)+(1 4 2 3)+(2 3)+(1 2)+(1 2 3)+(1 3 2)+(3 4 5)+(3 5)+(2 4 5 3)+(2 4 5)+(2 5 3)+(1 2)(4 5)+(1 2)(3 4 5)+(1 2)(3 5 4)+(1 2 3 4 5)+(1 2 4 5)+(1 2 5 4 3)+(1 2 5 3)+(1 2 5)+(1 2 5 3 4)+(1 2 5)(3 4)+(1 3 5 4 2)+(1 3)(4 5)+(1 3 5)+(1 3 2 4 5)+(1 3 5)(2 4)+(1 3)(2 5 4)+(1 3 2 5 4)+(1 3 4)(2 5)+(1 3 4 2 5)+(1 4 5 2)+(1 4 5 3)+(1 4 3 5)+(1 4 5 2 3)+(1 4)(2 3 5)+(1 4 3 2 5)+(1 4)(2 5)+(1 5 4 3 2)+(1 5 2)+(1 5 3)+(1 5 4)+(1 5)(3 4)+(1 5 2 3)+(1 5 3)(2 4)+(1 5 2 4 3)+(1 5 3 2 4)+(1 5)(2 4 3)+(1 5 2 4)$\\
\\$a_{12} = (2 4 3)+(2 4)+(1 2)(3 4)+(1 2 3 4)+(1 3 2 4)+(1 4 3 2)+(1 4 2)+(1 4 3)+(1 4 2 3)+(1 3)+(4 5)+(3 4 5)+(3 5 4)+(2 3)(4 5)+(2 3 4 5)+(2 3 5)+(2 4 5 3)+(2 4 5)+(2 4 3 5)+(2 5 3)+(2 5 3 4)+(2 5)(3 4)+(1 2)(3 5 4)+(1 2)(3 5)+(1 2 3 5 4)+(1 2 4 5)+(1 2 4)(3 5)+(1 2 5 3)+(1 2 5 3 4)+(1 3 2)(4 5)+(1 3 5 2)+(1 3)(4 5)+(1 3 5 4)+(1 3)(2 4 5)+(1 3 5 2 4)+(1 3 2 5 4)+(1 3 2 5)+(1 3 4)(2 5)+(1 4 5 3 2)+(1 4 2)(3 5)+(1 4 5 3)+(1 4 5)+(1 4 3 5)+(1 4 5)(2 3)+(1 4 3)(2 5)+(1 4)(2 5 3)+(1 4 3 2 5)+(1 4)(2 5)+(1 4 2 5)+(1 5 4 3 2)+(1 5 3 2)+(1 5 2)+(1 5 2)(3 4)+(1 5 3)+(1 5 4)+(1 5 2 3)+(1 5)(2 3)+(1 5 2 3 4)+(1 5 3)(2 4)+(1 5 3 2 4)+(1 5)(2 4 3)$\\
\\$a_{22} = (3 4)+(2 4)+(1 2 3 4)+(1 2 4 3)+(1 2 4)+(1 3 4 2)+(1 4 2)+(1 4 3)+(1 4)+(1 4 2 3)+(1 3 2)+(1 3)+(3 4 5)+(3 5 4)+(2 3)(4 5)+(2 3 5 4)+(2 4)(3 5)+(2 5 4 3)+(2 5 3)+(2 5)+(2 5 3 4)+(1 2)(4 5)+(1 2)(3 5 4)+(1 2)(3 5)+(1 2 3)(4 5)+(1 2 4)(3 5)+(1 2 4 3 5)+(1 2 5 4 3)+(1 2 5 3)+(1 2 5)+(1 3 4 5 2)+(1 3 5 4 2)+(1 3)(4 5)+(1 3 4 5)+(1 3 5 4)+(1 3 5)+(1 3)(2 4 5)+(1 3 2 4 5)+(1 3 5 2 4)+(1 3)(2 5 4)+(1 3)(2 5)+(1 3 4 2 5)+(1 4 5 2)+(1 4 3 5 2)+(1 4 5 3)+(1 4 5 2 3)+(1 4)(2 3 5)+(1 4 2 5 3)+(1 4)(2 5 3)+(1 4 3 2 5)+(1 4)(2 5)+(1 4 2 5)+(1 5 3 2)+(1 5 4 2)+(1 5 3 4 2)+(1 5 4 3)+(1 5)+(1 5 3 4)+(1 5)(2 3 4)+(1 5 3)(2 4)+(1 5 2 4 3)+(1 5 3 2 4)+(1 5)(2 4)$\\
\\$a_{32} = (2 3 4)+(1 2)(3 4)+(1 2 3 4)+(1 2 4)+(1 3 4 2)+(1 3 4)+(1 3 2 4)+(1 4 2)+(1 4 2 3)+(2 3)+(1 2)+(1 2 3)+(1 3 2)+(4 5)+(2 4 5 3)+(2 4)(3 5)+(2 4 3 5)+(2 5 4 3)+(2 5 4)+(2 5 3 4)+(1 2)(4 5)+(1 2 4 5)+(1 2 4)(3 5)+(1 2 5 3)+(1 2 5 3 4)+(1 3 5 2)+(1 3 4 5)+(1 3 5)+(1 3 2 4 5)+(1 3 5 2 4)+(1 3)(2 5 4)+(1 3)(2 5)+(1 3 2 5 4)+(1 4 2)(3 5)+(1 4 5 3)+(1 4 5)+(1 4 5 2 3)+(1 4)(2 3 5)+(1 4 3 2 5)+(1 4)(2 5)+(1 5 4 3 2)+(1 5 4 2)+(1 5 2)(3 4)+(1 5 4 3)+(1 5 3)+(1 5)+(1 5 3 4)+(1 5 4 2 3)+(1 5 4)(2 3)+(1 5)(2 3)+(1 5 3)(2 4)+(1 5 3 2 4)+(1 5)(2 4 3)+(1 5)(2 4)$\\
\\$a_{13} = \epsilon+(2 4 3)+(2 4)+(1 2)(3 4)+(1 2 3 4)+(1 2 4 3)+(1 2 4)+(1 3)(2 4)+(1 4 3 2)+(1 4 2)+(1 4)+(2 3)+(1 2)+(1 3 2)+(3 4 5)+(3 5)+(2 3)(4 5)+(2 3 4 5)+(2 4)(3 5)+(2 4 3 5)+(2 5 4 3)+(2 5 3)+(2 5 4)+(2 5 3 4)+(2 5)(3 4)+(1 2)(4 5)+(1 2)(3 5 4)+(1 2)(3 5)+(1 2 3 4 5)+(1 2 4 5 3)+(1 2 5 4 3)+(1 2 5)+(1 3 2)(4 5)+(1 3 5 2)+(1 3 4 5)+(1 3 5 4)+(1 3 5)+(1 3)(2 4 5)+(1 3 5 2 4)+(1 3 5)(2 4)+(1 3 2 5 4)+(1 3 2 5)+(1 3 4)(2 5)+(1 3 4 2 5)+(1 4 5 3 2)+(1 4 5 2)+(1 4 2)(3 5)+(1 4 3 5 2)+(1 4 5 3)+(1 4 5)+(1 4)(3 5)+(1 4 3 5)+(1 4 5)(2 3)+(1 4)(2 3 5)+(1 4 2 3 5)+(1 4 3)(2 5)+(1 4)(2 5 3)+(1 4 3 2 5)+(1 4 2 5)+(1 5 4 3 2)+(1 5 3 2)+(1 5 3 4 2)+(1 5 2)(3 4)+(1 5 4 3)+(1 5 4)+(1 5)+(1 5 3 4)+(1 5)(3 4)+(1 5 2 3)+(1 5)(2 3)+(1 5 2 3 4)+(1 5 2 4 3)+(1 5 2 4)$\\
\\$a_{23} = \epsilon+(2 4)+(1 2)(3 4)+(1 2 4)+(1 3 4)+(1 3)(2 4)+(1 3 2 4)+(1 4 3 2)+(1 4)(2 3)+(1 2)+(1 2 3)+(1 3 2)+(3 5 4)+(3 5)+(2 3)(4 5)+(2 3 4 5)+(2 3 5 4)+(2 4 5)+(2 4 3 5)+(2 5)+(2 5)(3 4)+(1 2)(3 4 5)+(1 2 4 5 3)+(1 2 4 5)+(1 2 4 3 5)+(1 2 5 4 3)+(1 2 5 3)+(1 2 5)+(1 2 5 3 4)+(1 3 4 5 2)+(1 3 5 4 2)+(1 3 5 2)+(1 3)(4 5)+(1 3 4 5)+(1 3 5)+(1 3 2 4 5)+(1 3)(2 5 4)+(1 3)(2 5)+(1 3 4)(2 5)+(1 3 4 2 5)+(1 4 5 3 2)+(1 4 5 2)+(1 4 3 5 2)+(1 4 5 3)+(1 4)(3 5)+(1 4 5 2 3)+(1 4 5)(2 3)+(1 4)(2 3 5)+(1 4 2 3 5)+(1 4 3)(2 5)+(1 4)(2 5 3)+(1 4 3 2 5)+(1 4)(2 5)+(1 5 3 2)+(1 5 2)+(1 5 3 4 2)+(1 5 2)(3 4)+(1 5 3)+(1 5 4)+(1 5)+(1 5 3 4)+(1 5 2 3)+(1 5 4)(2 3)+(1 5 3)(2 4)+(1 5)(2 4 3)+(1 5 2 4)+(1 5)(2 4)$\\
\\$a_{33} = \epsilon+(1 2 4 3)+(1 2 4)+(1 3 4 2)+(1 4 2)+(1 4 3)+(1 4 2 3)+(1 4)(2 3)+(1 2 3)+(1 3 2)+(3 4 5)+(3 5)+(2 3 4 5)+(2 4 5)+(2 4 3 5)+(2 5 3 4)+(1 2)(3 4 5)+(1 2 3)(4 5)+(1 2 3 5 4)+(1 2 4 5)+(1 2 5 4 3)+(1 2 5 3)+(1 2 5 4)+(1 2 5)+(1 2 5)(3 4)+(1 3 2)(4 5)+(1 3 5 4 2)+(1 3)(4 5)+(1 3)(2 4 5)+(1 3 5 2 4)+(1 3)(2 5 4)+(1 3)(2 5)+(1 3 2 5 4)+(1 3 2 5)+(1 3 4 2 5)+(1 4 5 2)+(1 4 3 5 2)+(1 4 5 3)+(1 4)(3 5)+(1 4 3 5)+(1 4 5 2 3)+(1 4 5)(2 3)+(1 4 2 5 3)+(1 4)(2 5 3)+(1 4 3 2 5)+(1 4)(2 5)+(1 4 2 5)+(1 5 3 2)+(1 5 4 2)+(1 5 2)+(1 5 4 3)+(1 5 3)+(1 5 4)+(1 5)(3 4)+(1 5 2 3)+(1 5 4)(2 3)+(1 5)(2 3 4)+(1 5 2 4 3)+(1 5)(2 4)$\\

Below are the entries for $M^a$:\\

\small
\noindent
$a_{11}= \epsilon+(3 4)+(2 3 4)+(2 4)+(1 2 4)+(1 3 4 2)+(1 3)(2 4)+(1 3 2 4)+(1 4 2)+(1 4 3)+(1 4)+(1 4)(2 3)+(2 3)+(1 2)+(4 5)+(3 4 5)+(2 4 5)+(2 4)(3 5)+(2 4 3 5)+(2 5 4 3)+(2 5)+(2 5)(3 4)+(1 2)(3 4 5)+(1 2)(3 5)+(1 2 3)(4 5)+(1 2 3 4 5)+(1 2 3 5 4)+(1 2 3 5)+(1 2 4 5 3)+(1 2 4 5)+(1 2 4)(3 5)+(1 2 4 3 5)+(1 2 5 4)+(1 2 5)+(1 3 2)(4 5)+(1 3 5 4 2)+(1 3)(4 5)+(1 3 5 4)+(1 3 5)+(1 3)(2 4 5)+(1 3 2 4 5)+(1 3 5 2 4)+(1 3 5)(2 4)+(1 3)(2 5 4)+(1 3)(2 5)+(1 3 2 5)+(1 3 4)(2 5)+(1 3 4 2 5)+(1 4 5 2)+(1 4 3 5 2)+(1 4 3 5)+(1 4 5)(2 3)+(1 4)(2 3 5)+(1 4 2 3 5)+(1 4 2 5 3)+(1 4)(2 5 3)+(1 4 3 2 5)+(1 4)(2 5)+(1 4 2 5)+(1 5 4 3 2)+(1 5 3 2)+(1 5 4 2)+(1 5 2)+(1 5 3 4 2)+(1 5 4)+(1 5)+(1 5 4 2 3)+(1 5 2 3)+(1 5 4)(2 3)+(1 5 3 2 4)+(1 5)(2 4 3)+(1 5 2 4)$\\
\\$a_{21}= (3 4)+(2 4 3)+(1 2 4)+(1 3 4 2)+(1 3 4)+(1 3)(2 4)+(1 3 2 4)+(1 4 3 2)+(1 4 3)+(1 4)(2 3)+(1 2)+(1 3)+(4 5)+(3 4 5)+(2 3)(4 5)+(2 3 4 5)+(2 3 5)+(2 4 5 3)+(2 4)(3 5)+(2 5 4 3)+(2 5 4)+(2 5)+(1 2 3 4 5)+(1 2 3 5)+(1 2 5 4 3)+(1 2 5 3)+(1 2 5 4)+(1 2 5 3 4)+(1 2 5)(3 4)+(1 3 2)(4 5)+(1 3 4 5 2)+(1 3 4 5)+(1 3 5)+(1 3)(2 4 5)+(1 3 5 2 4)+(1 3 5)(2 4)+(1 3)(2 5)+(1 3 4 2 5)+(1 4 5 3 2)+(1 4 3 5 2)+(1 4 5)+(1 4 5 2 3)+(1 4)(2 5 3)+(1 4 3 2 5)+(1 4 2 5)+(1 5 3 2)+(1 5 2)(3 4)+(1 5 4)+(1 5)(2 3)+(1 5 2 3 4)+(1 5 2 4 3)+(1 5)(2 4 3)+(1 5 2 4)$\\
\\$a_{31}= (2 3 4)+(2 4 3)+(1 2 4 3)+(1 3 4)+(1 4 2)+(1 4)+(1 4)(2 3)+(2 3)+(1 2 3)+(4 5)+(3 4 5)+(3 5 4)+(3 5)+(2 3)(4 5)+(2 3 5 4)+(2 4 5)+(2 4)(3 5)+(2 4 3 5)+(2 5 4 3)+(2 5 3)+(2 5 4)+(2 5 3 4)+(1 2)(4 5)+(1 2)(3 5 4)+(1 2)(3 5)+(1 2 3 5 4)+(1 2 3 5)+(1 2 4 5)+(1 3 4 5 2)+(1 3 5 4 2)+(1 3 5 2)+(1 3)(4 5)+(1 3 5)+(1 3)(2 4 5)+(1 3 2 4 5)+(1 3)(2 5 4)+(1 3 2 5 4)+(1 3 2 5)+(1 3 4 2 5)+(1 4 5 3 2)+(1 4 5 2)+(1 4 2)(3 5)+(1 4 3 5 2)+(1 4 5)+(1 4 3 5)+(1 4 5 2 3)+(1 4 2 3 5)+(1 4 3 2 5)+(1 4 2 5)+(1 5 2)(3 4)+(1 5 4)+(1 5)+(1 5 4 2 3)+(1 5 2 4 3)+(1 5 3 2 4)$\\
\\$a_{12}= (2 3 4)+(1 2 3 4)+(1 2 4 3)+(1 2 4)+(1 3 4 2)+(1 3)(2 4)+(1 4 3 2)+(1 4)(2 3)+(2 3)+(1 2 3)+(1 3)+(4 5)+(3 5 4)+(3 5)+(2 3 4 5)+(2 3 5 4)+(2 3 5)+(2 4 5 3)+(2 4)(3 5)+(2 5 3)+(2 5 4)+(2 5)+(2 5 3 4)+(1 2)(4 5)+(1 2)(3 4 5)+(1 2)(3 5)+(1 2 3)(4 5)+(1 2 3 5)+(1 2 4)(3 5)+(1 2 5 3)+(1 2 5 4)+(1 2 5 3 4)+(1 2 5)(3 4)+(1 3 2)(4 5)+(1 3 5 2)+(1 3)(4 5)+(1 3 4 5)+(1 3 5 4)+(1 3 2 4 5)+(1 3 5)(2 4)+(1 3)(2 5 4)+(1 3)(2 5)+(1 3 2 5 4)+(1 3 2 5)+(1 3 4)(2 5)+(1 4 5 2)+(1 4)(3 5)+(1 4 3 5)+(1 4 5)(2 3)+(1 4)(2 3 5)+(1 4 2 5 3)+(1 4)(2 5 3)+(1 4 3 2 5)+(1 4)(2 5)+(1 4 2 5)+(1 5 4 3 2)+(1 5 3 2)+(1 5 4 2)+(1 5 2)+(1 5 2)(3 4)+(1 5)+(1 5 2 3)+(1 5 4)(2 3)+(1 5)(2 3)+(1 5 3 2 4)+(1 5)(2 4 3)+(1 5)(2 4)$\\
\\$a_{22}= (3 4)+(2 4 3)+(1 2)(3 4)+(1 2 3 4)+(1 2 4 3)+(1 2 4)+(1 3 4)+(1 3 2 4)+(1 4)(2 3)+(1 2)+(1 2 3)+(1 3 2)+(1 3)+(3 4 5)+(2 3)(4 5)+(2 3 4 5)+(2 3 5 4)+(2 3 5)+(2 4 5 3)+(2 4 5)+(2 4)(3 5)+(2 5 3)+(2 5 3 4)+(2 5)(3 4)+(1 2)(3 4 5)+(1 2)(3 5 4)+(1 2 3)(4 5)+(1 2 3 5 4)+(1 2 3 5)+(1 2 4 5 3)+(1 2 4 5)+(1 2 5 4 3)+(1 2 5 4)+(1 3 4 5 2)+(1 3 5 4 2)+(1 3 5 2)+(1 3)(4 5)+(1 3 5)+(1 3 5 2 4)+(1 3)(2 5 4)+(1 3 2 5 4)+(1 3 2 5)+(1 3 4)(2 5)+(1 4 5 3 2)+(1 4 5 2)+(1 4 2)(3 5)+(1 4 5)+(1 4)(3 5)+(1 4)(2 3 5)+(1 4 2 5 3)+(1 4 3)(2 5)+(1 4)(2 5 3)+(1 4 3 2 5)+(1 5 4 3 2)+(1 5 3 2)+(1 5 4 2)+(1 5 2)(3 4)+(1 5)+(1 5 4 2 3)+(1 5 2 3)+(1 5 4)(2 3)+(1 5 2 4 3)+(1 5 3 2 4)+(1 5)(2 4 3)+(1 5)(2 4)$\\
\\$a_{32}= (3 4)+(2 4 3)+(2 4)+(1 2)(3 4)+(1 2 4)+(1 3 4 2)+(1 3 4)+(1 3 2 4)+(1 4 2)+(1 4 3)+(1 4 2 3)+(1 4)(2 3)+(2 3)+(1 3 2)+(4 5)+(3 4 5)+(3 5)+(2 3)(4 5)+(2 3 5)+(2 4 5)+(2 4)(3 5)+(2 4 3 5)+(2 5 4 3)+(2 5)+(2 5 3 4)+(2 5)(3 4)+(1 2)(3 5 4)+(1 2 3 4 5)+(1 2 3 5)+(1 2 4 3 5)+(1 2 5 4)+(1 3 2)(4 5)+(1 3 5 2)+(1 3)(4 5)+(1 3 4 5)+(1 3 5)+(1 3)(2 4 5)+(1 3 2 4 5)+(1 3)(2 5 4)+(1 3)(2 5)+(1 3 2 5 4)+(1 3 4)(2 5)+(1 3 4 2 5)+(1 4 5 2)+(1 4 3 5 2)+(1 4 5 3)+(1 4)(3 5)+(1 4 3 5)+(1 4 5)(2 3)+(1 4)(2 3 5)+(1 4 2 5 3)+(1 4 3)(2 5)+(1 4)(2 5 3)+(1 4)(2 5)+(1 4 2 5)+(1 5 3 2)+(1 5 2)+(1 5 3 4 2)+(1 5 2)(3 4)+(1 5 4 3)+(1 5)(3 4)+(1 5 2 3)+(1 5 4)(2 3)+(1 5)(2 3)+(1 5 3)(2 4)+(1 5 2 4 3)+(1 5 3 2 4)+(1 5 2 4)$\\
\\$a_{13}= \epsilon+(3 4)+(1 2 4 3)+(1 3 4 2)+(1 3 2 4)+(1 4)+(1 4 2 3)+(1 4)(2 3)+(1 2)+(1 3 2)+(4 5)+(3 5 4)+(3 5)+(2 3)(4 5)+(2 3 4 5)+(2 4)(3 5)+(2 5 4 3)+(2 5 4)+(2 5 3 4)+(1 2)(4 5)+(1 2)(3 5)+(1 2 3 5 4)+(1 2 3 5)+(1 2 5)(3 4)+(1 3 4 5)+(1 3 5 2 4)+(1 3 5)(2 4)+(1 3 2 5 4)+(1 3 4)(2 5)+(1 4 5 3 2)+(1 4 5 3)+(1 4 5 2 3)+(1 4 3)(2 5)+(1 4)(2 5)+(1 5 4 3 2)+(1 5 3 2)+(1 5 2)+(1 5 3 4 2)+(1 5 3)+(1 5)+(1 5 4 2 3)+(1 5 3)(2 4)+(1 5)(2 4 3)$\\
\\$a_{23}= \epsilon+(3 4)+(2 4)+(1 2 4 3)+(1 2 4)+(1 4)+(2 3)+(1 2 3)+(1 3 2)+(1 3)+(4 5)+(3 5 4)+(2 3)(4 5)+(2 4 5 3)+(2 4 5)+(2 4)(3 5)+(2 4 3 5)+(2 5 4 3)+(2 5 3)+(2 5 4)+(2 5)+(2 5)(3 4)+(1 2)(3 5)+(1 2 3 4 5)+(1 2 4 5 3)+(1 2 4)(3 5)+(1 2 5 4)+(1 2 5)+(1 3 2)(4 5)+(1 3 5 4 2)+(1 3 5 2)+(1 3)(4 5)+(1 3 4 5)+(1 3 5 4)+(1 3 5)+(1 3)(2 4 5)+(1 3 5 2 4)+(1 3 5)(2 4)+(1 3 2 5)+(1 3 4 2 5)+(1 4 2)(3 5)+(1 4 5)+(1 4)(3 5)+(1 4 3 5)+(1 4 5)(2 3)+(1 4)(2 3 5)+(1 4 2 5 3)+(1 4)(2 5 3)+(1 4)(2 5)+(1 4 2 5)+(1 5 4 2)+(1 5 2)+(1 5 2)(3 4)+(1 5 3)+(1 5)(3 4)+(1 5 2 3)+(1 5 4)(2 3)+(1 5 2 3 4)+(1 5 3 2 4)+(1 5 2 4)+(1 5)(2 4)$\\
\\$a_{33}= (1 2)(3 4)+(1 2 3 4)+(1 2 4 3)+(1 3 4 2)+(1 3 4)+(1 3)(2 4)+(1 3 2 4)+(1 4 3)+(1 4)+(1 4 2 3)+(2 3)+(1 2)+(1 2 3)+(1 3)+(3 5 4)+(3 5)+(2 3)(4 5)+(2 3 5)+(2 4 3 5)+(2 5 3 4)+(2 5)(3 4)+(1 2 3 4 5)+(1 2 3 5)+(1 2 4)(3 5)+(1 2 4 3 5)+(1 2 5 3)+(1 2 5 4)+(1 2 5 3 4)+(1 2 5)(3 4)+(1 3 4 5 2)+(1 3 5 4 2)+(1 3 5 2)+(1 3 4 5)+(1 3 5 4)+(1 3 5)+(1 3)(2 4 5)+(1 3 2 4 5)+(1 3 5)(2 4)+(1 3)(2 5)+(1 4 2)(3 5)+(1 4 3 5 2)+(1 4 3 5)+(1 4)(2 3 5)+(1 4 2 5 3)+(1 4 3)(2 5)+(1 4)(2 5 3)+(1 4 3 2 5)+(1 4)(2 5)+(1 4 2 5)+(1 5 3 2)+(1 5 3 4 2)+(1 5 4 3)+(1 5 4)+(1 5)+(1 5 3 4)+(1 5)(3 4)+(1 5 4)(2 3)+(1 5)(2 3 4)+(1 5 3 2 4)+(1 5)(2 4 3)+(1 5)(2 4)$\\

Below are the entries for $M^b$:\\

\small
\noindent
$a_{11}=\epsilon+(3 4)+(2 3 4)+(2 4)+(1 2 4)+(1 3 4 2)+(1 3)(2 4)+(1 3 2 4)+(1 4 2)+(1 4 3)+(1 4)+(1 4)(2 3)+(2 3)+(1 2)+(4 5)+(3 4 5)+(2 4 5)+(2 4)(3 5)+(2 4 3 5)+(2 5 4 3)+(2 5)+(2 5)(3 4)+(1 2)(3 4 5)+(1 2)(3 5)+(1 2 3)(4 5)+(1 2 3 4 5)+(1 2 3 5 4)+(1 2 3 5)+(1 2 4 5 3)+(1 2 4 5)+(1 2 4)(3 5)+(1 2 4 3 5)+(1 2 5 4)+(1 2 5)+(1 3 2)(4 5)+(1 3 5 4 2)+(1 3)(4 5)+(1 3 5 4)+(1 3 5)+(1 3)(2 4 5)+(1 3 2 4 5)+(1 3 5 2 4)+(1 3 5)(2 4)+(1 3)(2 5 4)+(1 3)(2 5)+(1 3 2 5)+(1 3 4)(2 5)+(1 3 4 2 5)+(1 4 5 2)+(1 4 3 5 2)+(1 4 3 5)+(1 4 5)(2 3)+(1 4)(2 3 5)+(1 4 2 3 5)+(1 4 2 5 3)+(1 4)(2 5 3)+(1 4 3 2 5)+(1 4)(2 5)+(1 4 2 5)+(1 5 4 3 2)+(1 5 3 2)+(1 5 4 2)+(1 5 2)+(1 5 3 4 2)+(1 5 4)+(1 5)+(1 5 4 2 3)+(1 5 2 3)+(1 5 4)(2 3)+(1 5 3 2 4)+(1 5)(2 4 3)+(1 5 2 4)$\\
\\$a_{21}=(3 4)+(2 4 3)+(1 2 4)+(1 3 4 2)+(1 3 4)+(1 3)(2 4)+(1 3 2 4)+(1 4 3 2)+(1 4 3)+(1 4)(2 3)+(1 2)+(1 3)+(4 5)+(3 4 5)+(2 3)(4 5)+(2 3 4 5)+(2 3 5)+(2 4 5 3)+(2 4)(3 5)+(2 5 4 3)+(2 5 4)+(2 5)+(1 2 3 4 5)+(1 2 3 5)+(1 2 5 4 3)+(1 2 5 3)+(1 2 5 4)+(1 2 5 3 4)+(1 2 5)(3 4)+(1 3 2)(4 5)+(1 3 4 5 2)+(1 3 4 5)+(1 3 5)+(1 3)(2 4 5)+(1 3 5 2 4)+(1 3 5)(2 4)+(1 3)(2 5)+(1 3 4 2 5)+(1 4 5 3 2)+(1 4 3 5 2)+(1 4 5)+(1 4 5 2 3)+(1 4)(2 5 3)+(1 4 3 2 5)+(1 4 2 5)+(1 5 3 2)+(1 5 2)(3 4)+(1 5 4)+(1 5)(2 3)+(1 5 2 3 4)+(1 5 2 4 3)+(1 5)(2 4 3)+(1 5 2 4)$\\
\\$a_{31}=(2 3 4)+(2 4 3)+(1 2 4 3)+(1 3 4)+(1 4 2)+(1 4)+(1 4)(2 3)+(2 3)+(1 2 3)+(4 5)+(3 4 5)+(3 5 4)+(3 5)+(2 3)(4 5)+(2 3 5 4)+(2 4 5)+(2 4)(3 5)+(2 4 3 5)+(2 5 4 3)+(2 5 3)+(2 5 4)+(2 5 3 4)+(1 2)(4 5)+(1 2)(3 5 4)+(1 2)(3 5)+(1 2 3 5 4)+(1 2 3 5)+(1 2 4 5)+(1 3 4 5 2)+(1 3 5 4 2)+(1 3 5 2)+(1 3)(4 5)+(1 3 5)+(1 3)(2 4 5)+(1 3 2 4 5)+(1 3)(2 5 4)+(1 3 2 5 4)+(1 3 2 5)+(1 3 4 2 5)+(1 4 5 3 2)+(1 4 5 2)+(1 4 2)(3 5)+(1 4 3 5 2)+(1 4 5)+(1 4 3 5)+(1 4 5 2 3)+(1 4 2 3 5)+(1 4 3 2 5)+(1 4 2 5)+(1 5 2)(3 4)+(1 5 4)+(1 5)+(1 5 4 2 3)+(1 5 2 4 3)+(1 5 3 2 4)$\\
\\$a_{12}=(2 3 4)+(1 2 3 4)+(1 2 4 3)+(1 2 4)+(1 3 4 2)+(1 3)(2 4)+(1 4 3 2)+(1 4)(2 3)+(2 3)+(1 2 3)+(1 3)+(4 5)+(3 5 4)+(3 5)+(2 3 4 5)+(2 3 5 4)+(2 3 5)+(2 4 5 3)+(2 4)(3 5)+(2 5 3)+(2 5 4)+(2 5)+(2 5 3 4)+(1 2)(4 5)+(1 2)(3 4 5)+(1 2)(3 5)+(1 2 3)(4 5)+(1 2 3 5)+(1 2 4)(3 5)+(1 2 5 3)+(1 2 5 4)+(1 2 5 3 4)+(1 2 5)(3 4)+(1 3 2)(4 5)+(1 3 5 2)+(1 3)(4 5)+(1 3 4 5)+(1 3 5 4)+(1 3 2 4 5)+(1 3 5)(2 4)+(1 3)(2 5 4)+(1 3)(2 5)+(1 3 2 5 4)+(1 3 2 5)+(1 3 4)(2 5)+(1 4 5 2)+(1 4)(3 5)+(1 4 3 5)+(1 4 5)(2 3)+(1 4)(2 3 5)+(1 4 2 5 3)+(1 4)(2 5 3)+(1 4 3 2 5)+(1 4)(2 5)+(1 4 2 5)+(1 5 4 3 2)+(1 5 3 2)+(1 5 4 2)+(1 5 2)+(1 5 2)(3 4)+(1 5)+(1 5 2 3)+(1 5 4)(2 3)+(1 5)(2 3)+(1 5 3 2 4)+(1 5)(2 4 3)+(1 5)(2 4)$\\
\\$a_{22}=(3 4)+(2 4 3)+(1 2)(3 4)+(1 2 3 4)+(1 2 4 3)+(1 2 4)+(1 3 4)+(1 3 2 4)+(1 4)(2 3)+(1 2)+(1 2 3)+(1 3 2)+(1 3)+(3 4 5)+(2 3)(4 5)+(2 3 4 5)+(2 3 5 4)+(2 3 5)+(2 4 5 3)+(2 4 5)+(2 4)(3 5)+(2 5 3)+(2 5 3 4)+(2 5)(3 4)+(1 2)(3 4 5)+(1 2)(3 5 4)+(1 2 3)(4 5)+(1 2 3 5 4)+(1 2 3 5)+(1 2 4 5 3)+(1 2 4 5)+(1 2 5 4 3)+(1 2 5 4)+(1 3 4 5 2)+(1 3 5 4 2)+(1 3 5 2)+(1 3)(4 5)+(1 3 5)+(1 3 5 2 4)+(1 3)(2 5 4)+(1 3 2 5 4)+(1 3 2 5)+(1 3 4)(2 5)+(1 4 5 3 2)+(1 4 5 2)+(1 4 2)(3 5)+(1 4 5)+(1 4)(3 5)+(1 4)(2 3 5)+(1 4 2 5 3)+(1 4 3)(2 5)+(1 4)(2 5 3)+(1 4 3 2 5)+(1 5 4 3 2)+(1 5 3 2)+(1 5 4 2)+(1 5 2)(3 4)+(1 5)+(1 5 4 2 3)+(1 5 2 3)+(1 5 4)(2 3)+(1 5 2 4 3)+(1 5 3 2 4)+(1 5)(2 4 3)+(1 5)(2 4)$\\
\\$a_{32}=(3 4)+(2 4 3)+(2 4)+(1 2)(3 4)+(1 2 4)+(1 3 4 2)+(1 3 4)+(1 3 2 4)+(1 4 2)+(1 4 3)+(1 4 2 3)+(1 4)(2 3)+(2 3)+(1 3 2)+(4 5)+(3 4 5)+(3 5)+(2 3)(4 5)+(2 3 5)+(2 4 5)+(2 4)(3 5)+(2 4 3 5)+(2 5 4 3)+(2 5)+(2 5 3 4)+(2 5)(3 4)+(1 2)(3 5 4)+(1 2 3 4 5)+(1 2 3 5)+(1 2 4 3 5)+(1 2 5 4)+(1 3 2)(4 5)+(1 3 5 2)+(1 3)(4 5)+(1 3 4 5)+(1 3 5)+(1 3)(2 4 5)+(1 3 2 4 5)+(1 3)(2 5 4)+(1 3)(2 5)+(1 3 2 5 4)+(1 3 4)(2 5)+(1 3 4 2 5)+(1 4 5 2)+(1 4 3 5 2)+(1 4 5 3)+(1 4)(3 5)+(1 4 3 5)+(1 4 5)(2 3)+(1 4)(2 3 5)+(1 4 2 5 3)+(1 4 3)(2 5)+(1 4)(2 5 3)+(1 4)(2 5)+(1 4 2 5)+(1 5 3 2)+(1 5 2)+(1 5 3 4 2)+(1 5 2)(3 4)+(1 5 4 3)+(1 5)(3 4)+(1 5 2 3)+(1 5 4)(2 3)+(1 5)(2 3)+(1 5 3)(2 4)+(1 5 2 4 3)+(1 5 3 2 4)+(1 5 2 4)$\\
\\$a_{13}=\epsilon+(3 4)+(1 2 4 3)+(1 3 4 2)+(1 3 2 4)+(1 4)+(1 4 2 3)+(1 4)(2 3)+(1 2)+(1 3 2)+(4 5)+(3 5 4)+(3 5)+(2 3)(4 5)+(2 3 4 5)+(2 4)(3 5)+(2 5 4 3)+(2 5 4)+(2 5 3 4)+(1 2)(4 5)+(1 2)(3 5)+(1 2 3 5 4)+(1 2 3 5)+(1 2 5)(3 4)+(1 3 4 5)+(1 3 5 2 4)+(1 3 5)(2 4)+(1 3 2 5 4)+(1 3 4)(2 5)+(1 4 5 3 2)+(1 4 5 3)+(1 4 5 2 3)+(1 4 3)(2 5)+(1 4)(2 5)+(1 5 4 3 2)+(1 5 3 2)+(1 5 2)+(1 5 3 4 2)+(1 5 3)+(1 5)+(1 5 4 2 3)+(1 5 3)(2 4)+(1 5)(2 4 3)$\\
\\$a_{23}=\epsilon+(3 4)+(2 4)+(1 2 4 3)+(1 2 4)+(1 4)+(2 3)+(1 2 3)+(1 3 2)+(1 3)+(4 5)+(3 5 4)+(2 3)(4 5)+(2 4 5 3)+(2 4 5)+(2 4)(3 5)+(2 4 3 5)+(2 5 4 3)+(2 5 3)+(2 5 4)+(2 5)+(2 5)(3 4)+(1 2)(3 5)+(1 2 3 4 5)+(1 2 4 5 3)+(1 2 4)(3 5)+(1 2 5 4)+(1 2 5)+(1 3 2)(4 5)+(1 3 5 4 2)+(1 3 5 2)+(1 3)(4 5)+(1 3 4 5)+(1 3 5 4)+(1 3 5)+(1 3)(2 4 5)+(1 3 5 2 4)+(1 3 5)(2 4)+(1 3 2 5)+(1 3 4 2 5)+(1 4 2)(3 5)+(1 4 5)+(1 4)(3 5)+(1 4 3 5)+(1 4 5)(2 3)+(1 4)(2 3 5)+(1 4 2 5 3)+(1 4)(2 5 3)+(1 4)(2 5)+(1 4 2 5)+(1 5 4 2)+(1 5 2)+(1 5 2)(3 4)+(1 5 3)+(1 5)(3 4)+(1 5 2 3)+(1 5 4)(2 3)+(1 5 2 3 4)+(1 5 3 2 4)+(1 5 2 4)+(1 5)(2 4)$\\
\\$a_{33}=(1 2)(3 4)+(1 2 3 4)+(1 2 4 3)+(1 3 4 2)+(1 3 4)+(1 3)(2 4)+(1 3 2 4)+(1 4 3)+(1 4)+(1 4 2 3)+(2 3)+(1 2)+(1 2 3)+(1 3)+(3 5 4)+(3 5)+(2 3)(4 5)+(2 3 5)+(2 4 3 5)+(2 5 3 4)+(2 5)(3 4)+(1 2 3 4 5)+(1 2 3 5)+(1 2 4)(3 5)+(1 2 4 3 5)+(1 2 5 3)+(1 2 5 4)+(1 2 5 3 4)+(1 2 5)(3 4)+(1 3 4 5 2)+(1 3 5 4 2)+(1 3 5 2)+(1 3 4 5)+(1 3 5 4)+(1 3 5)+(1 3)(2 4 5)+(1 3 2 4 5)+(1 3 5)(2 4)+(1 3)(2 5)+(1 4 2)(3 5)+(1 4 3 5 2)+(1 4 3 5)+(1 4)(2 3 5)+(1 4 2 5 3)+(1 4 3)(2 5)+(1 4)(2 5 3)+(1 4 3 2 5)+(1 4)(2 5)+(1 4 2 5)+(1 5 3 2)+(1 5 3 4 2)+(1 5 4 3)+(1 5 4)+(1 5)+(1 5 3 4)+(1 5)(3 4)+(1 5 4)(2 3)+(1 5)(2 3 4)+(1 5 3 2 4)+(1 5)(2 4 3)+(1 5)(2 4)$\\

\end{landscape}

\end{document}